\documentclass[a4paper,fleqn,usenatbib]{mnras}

\usepackage{newtxtext,newtxmath}
\usepackage[T1]{fontenc}
\usepackage{ae,aecompl}

\usepackage{float}

\usepackage{graphicx}	% Including figure files
\usepackage{amsmath}	% Advanced maths commands
\usepackage{amssymb}	% Extra maths symbols
\usepackage{subfigure}
\newcommand{\kms}{km~s$^{-1}$}

\newcommand{\Mearth}{M$_\oplus$}
\title[]{ALMA observations of the multiplanet system 61~Vir: What lies
  outside super-Earth systems?}

\author[]{S. Marino$^{1}$\thanks{E-mail: s.marino@ast.cam.ac.uk}, M. C. Wyatt$^{1}$, G. M. Kennedy$^{1}$, W. Holland$^{2}$, L. Matr\`{a}$^{1}$, A. Shannon$^{3,4}$
  \newauthor{and R. J. Ivison$^{5,6}$.} \\
  $^{1}$Institute of Astronomy, University of Cambridge, Madingley Road, Cambridge CB3 0HA, UK\\
  $^{2}$UK Astronomy Technology Centre, Royal Observatory, Blackford Hill, Edinburgh EH9 3HJ, UK\\
  $^{3}$Department of Astronomy \& Astrophysics, The Pennsylvania State University, State College, PA 16801, USA\\
  $^{4}$Center for Exoplanets and Habitable Worlds, The Pennsylvania State University, State College, PA 16802, USA\\
  $^{5}$Institute for Astronomy, University of Edinburgh, Royal Observatory, Edinburgh EH9 3HJ, UK\\
  $^{6}$European Southern Observatory, Karl-Schwarzchild-Str 2, D-95748 Garching b. Munchen, Germany\\
}

\begin{document}
\label{firstpage}
\pagerange{\pageref{firstpage}--\pageref{lastpage}}
\maketitle

\begin{abstract}
  A decade of surveys has hinted at a possible higher occurrence rate
  of debris discs in systems hosting low mass planets. This could be
  due to common favourable forming conditions for rocky planets close
  in and planetesimals at large radii. In this paper we present the
  first resolved millimetre study of the debris disc in the 4.6 Gyr
  old multiplanet system 61~Vir, combining ALMA and JCMT data at 0.86
  mm. We fit the data using a parametric disc model, finding that the
  disc of planetesimals extends from 30 AU to at least 150 AU, with a
  surface density distribution of millimetre sized grains with a power
  law slope of $0.1^{+1.1}_{-0.8}$. We also present a numerical
  collisional model that can predict the evolution of the surface
  density of millimetre grains for a given primordial disc, finding
  that it does not necessarily have the same radial profile as the
  total mass surface density (as previous studies suggested for the
  optical depth), with the former being flatter. Finally, we find
  that if the planetesimal disc was stirred at 150 AU by an additional
  unseen planet, that planet should be more massive than 10 M$_\oplus$
  and lie between 10-20~AU. Lower planet masses and semi-major axes
  down to 4 AU are possible for eccentricities $\gg0.1$.
  \medskip
  %% \\
%% it should be a single paragraph not more than 250 words (200
%%   words for Letters).  No references should appear in the abstract.
\end{abstract}


%%%%%%%%%%%%%%%%%%%%%%%%%%%%%%%%%%%%%%%%%%%%%%%%%%

%%%%%%%%%%%%%%%%% BODY OF PAPER %%%%%%%%%%%%%%%%%%

\section{Introduction}
\label{sec:intro}

%% \textcolor{red}{Kuiper belt analogues are common around FGK stars}.
%% \\

Planetary systems around main sequence stars are not only composed of
planets; planetesimal belts can be also present, analogous to the
Kuiper belt (at tens of AU) and the Asteroid belt (within a few AU) in
the Solar System. These belts can produce dusty debris discs as the
result of a so-called collisional cascade \citep[e.g.,
][]{Dominik2003, Wyatt2007collisionalcascade}, where solids in a wide
size distribution from $\mu$m-sized grains up to km-sized
planetesimals are ground down in collisions, sustaining high levels of
dust and infrared excess over Gyr timescales. Debris discs at tens of
AU are fairly common around FGK stars, with occurrence rates of at
least $\sim20\%$ \citep[e.g.,][]{Su2006, Hillenbrand2008,
  Carpenter2009, Eiroa2013, Thureau2014, Matthews2014pp6,
  Montesinos2016}; therefore, a complete understanding of their
properties can give us information about planet formation and
migration history of planets in these systems
\citep[e.g.,][]{Wyatt2006}.

Moreover, a few systems are known to host both a planet(s) and a
debris disc(s). Among the best studied are $\beta$~Pic
\citep[e.g.,][]{Smith1984betapic, Lagrange2009betapicb, Dent2014},
HR~8799 \citep[e.g.,][]{Marois2008, Marois2010, Matthews2014hr8799,
  Booth2016} and Fomalhaut \citep[e.g.,][]{Kalas2008}, all with
planets directly imaged and lying between the star and the
disc. However, these systems are outliers in terms of their planets
and disc properties and neither represents the bulk of the known
planetary systems, nor debris discs.

Thanks to unbiased debris disc surveys of FGK stars within 45 pc
\citep[e.g., DUNES and DEBRIS][]{Eiroa2013, Matthews2014pp6}, it has
been possible to study the frequency of circumstellar material around
stars hosting high- and low-mass planets detected by radial velocity
surveys. Studies focused on high-mass planets found no evidence of a
different debris disc incidence rate in these planet hosting stars
compared to normal field stars \citep{Greaves2004, Moro-Martin2007,
  Bryden2009}. On the other hand, two studies focused on planetary
systems with planet masses below $\sim95$~\Mearth found debris disc
incidence rates of: 4/6 \citep{Wyatt2012}, significantly higher
compared to field stars; and 2/6 \citep{Moro-Martin2015}, consistent
with field stars. Combining both samples, at least four out of eight
systems with low-mass planets also have a debris disc, which suggests
that there might be a difference in the occurrence of bright debris
discs in systems with low-mass planets, as predicted by planet
formation models \cite[e.g.,][]{Raymond2011}.

%% Although in the latter which focused on far-IR disc
%% observations, no excess was detected in the planet-hosting star
%% HD~69830, while it is known to host a debris disc from previous mid-IR
%% observations. 

One of these planetary systems hosting a debris disc is 61~Vir. This
system located at 8.6 pc \citep{vanLeeuwen2007} is composed of : 1) a
G5 $4.6\pm0.9$ Gyr old star \citep{Wright2011, Vican2012}; 2) three RV
planets of minimum masses 5, 18 and 23~M$_\oplus$ and semi-major axes
of 0.05, 0.22 and 0.49 AU, respectively (\citealt{Vogt2010}, the third
one was not confirmed in the HARPS data, \citealt{Wyatt2012}; and 3) a
debris disc discovered by Spitzer with a fractional luminosity $L_{\rm
  disc}/L_\star$ of $2\times10^{-5}$ \citep{Bryden2006}. The disc was
later imaged by Herschel showing that the disc density peaks between
30 and 100 AU and it is inclined by $\sim77\degr$ with respect to the
plane of the sky \citep{Wyatt2012}. If disc and orbits of these
planets are co-planar, then the planet masses would be only
underestimated by 3\%. This system is particularly interesting as the
fraction of stars with super-Earths, similar to 61 Vir, could be up to
30-50\% \citep[e.g.,][]{Howard2010, Mayor2011, Fressin2013}, which
makes 61~Vir a good case to study the formation of such abundant
planets by analysing its debris disc.

Due to a low 50 AU resolution, Herschel could not constrain the exact
morphology and dust distribution at the inner regions of the disc, but
by image and spectral energy distribution (SED) fitting
\cite{Wyatt2012} found three best fit models: 1) an extended disc with
a sharp inner edge at $\sim$30 AU, extending at least out to 100~AU,
and a surface density or optical depth radial profile with an exponent
of -1.1; 2) similar to the first model, but adding an inner component
where the surface density increases with radius as $r^{7/3}$ (inspired
by collisional evolution models) from 1 AU to the disc inner edge now
placed at 43 AU; 3) a two belt model consisting of two 10 AU wide
dusty belts centered at 40 and 90 AU. These three models could well
fit the previous observations, but the low 50 AU resolution hindered
determining the exact dust distribution. Moreover, because the disc
emission at Herschel wavelengths is dominated by small grains that are
subject to radiation forces, the derived distribution does not
necessarily trace the location of the parent planetesimal belt, as
they can extend to larger radii beyond that belt
\citep[e.g.,][]{Thebault2007}.

%% disc model with a sharp inner edge, or peak in the optical depth at 40
%% AU with a model that is collisionally depleted from 1 to 40 AU,
%% i.e. $\tau\propto r^{7/3}$. On the other hand, the outer edge was only
%% constrained to be greater than 100 AU. Moreover, at Herschel
%% wavelengths the disc emission is dominated by small grains that are
%% subject to radiation forces, therefore, they do not directly trace the
%% location of the parent planetesimal belt as they can extend up to
%% larger radii.

%% Although its disc was marginally resolved with a 50 AU resultion
%% between 60-500 $\mu$

%% understanding to study the interaction and formatio understand
%% planet formation because of the high prevalecence of super-Earths and
%% the pressence of a debris disc.

%% This is the only system that we know with
%% super-Earths and a debris disc, which makes it particularly
%% interesting as the fraction of stars with such This system with
%% super-Earths represents one of the most common planetary systems t

%% , then their masses have orbits These type of
%% systems with super-Earth represents 

%% \textcolor{red}{Super
%%   earths and why 61 Vir is important}.  \\

In this paper we present the first observations of 61~Vir with the
Atacama Large Millimeter/submillimeter Array (ALMA) at 0.86~mm,
obtained with the aim of studying its debris disc to reveal the
location of the parent planetesimals, and place constraints on the
presence of planets at large separations that can shape the mass
distribution in the disc. Because radiation forces are negligible for
mm-sized grains, their distribution can be used to trace the location
of the biggest km-sized planetesimals (or bigger), which contain the
bulk of the disc mass and sustain the collisional cascade. At
millimetre wavelengths, the dust thermal emission is dominated by
mm-sized grains ($\sim0.1-10$ mm), therefore, observations with ALMA
are well suited to study the dynamics and origin of debris discs. In
order to obtain the best disc constraints, in our analysis we combine
new ALMA band 7 observations and new data at 0.85 mm from the
Sub-millimetre Common-User Bolometer Array 2 (SCUBA2) installed in the
James Clerk Maxwell Telescope (JCMT), thus, incorporating information
from small and large angular scale structure.

In addition, we implement a simple numerical collisional evolution
model that simulates the evolution of a broad disc, taking into
account the disruption threshold of planetesimals as a function of
size, how relative velocities vary with radii, and the different
features in the size distribution of solids, e.g. the ripples close to
the blow-out size. We use this to constrain the initial solid mass or
surface density in the disc and the maximum planetesimal size.

%% Because small grains are subject to radiation pressure
%% ($\lesssim$100 $\mu$m-sized grains), observations in the optical,
%% NIR or MIR trace a mid-infrared wavelengths experienced destructive
%% collisions Because small grains are subject to radiation pressure,

This paper is organised as follows. In Sec. \ref{sec:scuba} we present
new SCUBA2/JCMT data. In Sec. \ref{sec:alma} we describe the ALMA
observations, studying the dust continuum and how it compares with
previous Herschel observations. We also search for CO v=0, J=3-2) line
emission. In Sec. \ref{sec:model} we fit the SCUBA and ALMA data using
a parametric disc model to study the distribution of millimetre dust
in the disc. Sec. \ref{sec:collmodel} describes a numerical model to
calculate the collisional evolution of a disc at different radii that
can be used to compare with observations determining the maximum
planetesimal size in a disc and the initial sold mass. In
Sec. \ref{discussion} we discuss the observations and possible
scenarios than could explain the low initial solid mass and maximum
planetesimal size. Moreover, we constrain the mass, semi-major axis
and eccentricity of a hypothetical planet stirring the disc. Finally,
In Sec. \ref{conclusions} we summarise and present the main
conclusions of this paper.

\section{SCUBA2 observations}
\label{sec:scuba}

As part of the SCUBA-2 Observations of Nearby Stars (SONS) survey
\citep{Panic2013}, 61~Vir was observed at 0.85 mm with SCUBA2/JCMT
\citep{Holland2013} to constrain the millimetre flux and extent of its
debris disc. 61 Vir was observed for 7.5 h and the data was reduced
using the Dynamic Iterative Map-Maker within the Starlink
\textsc{SMURF} package \citep{Chapin2013}, which was called from the
automated pipeline ORAC-DR \citep{Cavanagh2008}. More details on the
SCUBA2 data reduction of the SONS survey can be found in
\cite{Matthews2015} and \cite{Kennedy2015superearths}.

%% We re-reduce the SCUBA2
%% 0.85 mm data presented in \cite{Panic2013}. \textcolor{red}{Words
%%   about the re-reduction of the data by Wayne}.

\begin{figure}
    
  \includegraphics[trim=0.0cm 0.5cm 0.0cm 1.0cm, clip=true,
    width=1.0\columnwidth]{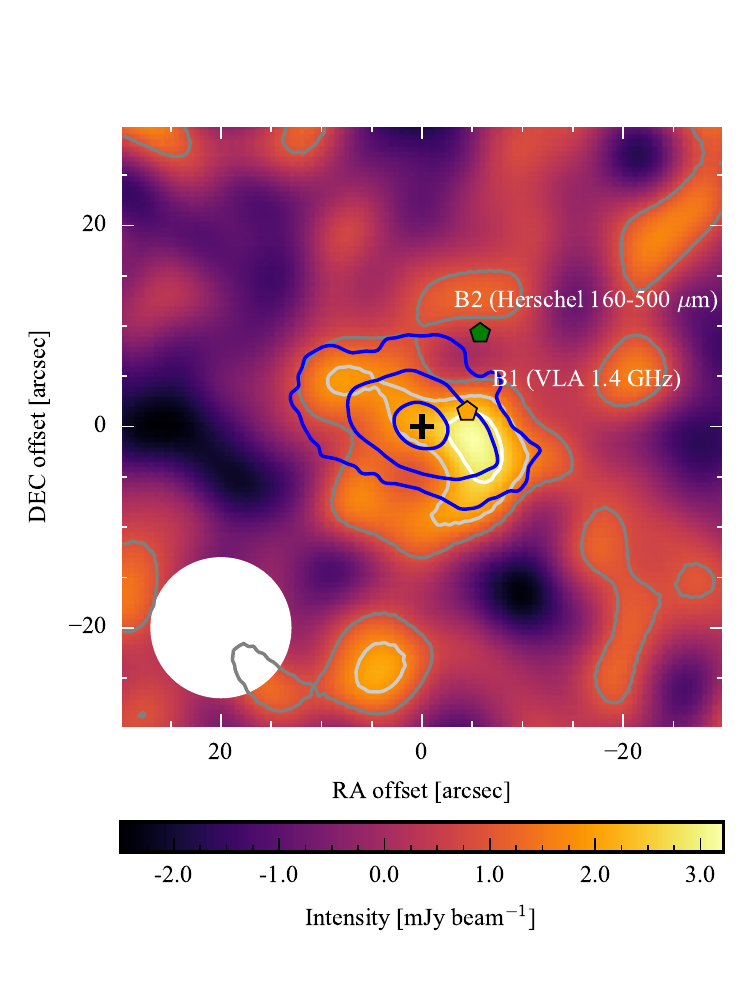}

  \caption{SCUBA2 0.85 mm continuum image of 61~Vir after subtracting
    two point sources from background emission. The beam size is
    $13\arcsec$ and is represented with a white ellipse at the bottom
    left corner of the image. The grey and white contours represent
    emission above 1, 2 and 3 times the noise level. Blue contours at
    arbitrary levels from the Herschel 70 $\mu$m image are overlayed
    and are corrected for proper motion. The green and yellow pentagon
    symbols indicate the position of the background sources that were
    subtracted from this image. The x- and y-axes indicate the offset
    from the stellar position in R.A. and decl. in arcsec, i.e. north
    is up and east is left. The stellar position is marked with a
    black ``+''.}
  \label{fig:scuba}
  
\end{figure}

Herschel and VLA observations previously found three background
sources close to 61~Vir which could affect our analysis of the SCUBA2
data. To obtain a non-contaminated large scale image and photometry of
61~Vir we subtract these as point sources, using as PSF the SCUBA2
reduced observation of Uranus obtained in the same run. As two of
these sources are detected in the ALMA data (see
Sec. \ref{sec:continuum}), we can derive their fluxes and astrometric
positions at 0.86~mm, and accurately subtract these from the SCUBA2
reduced image, correcting for the proper motion of 61~Vir $\mu$=(1.07,
-1.06) $\arcsec$~yr$^{-1}$ \citep{vanLeeuwen2007}. The third
background and more distant source from 61 Vir is not detected with
SCUBA2, and lies outside the field of view of the ALMA
observations. In Figure \ref{fig:scuba} we present the SCUBA2 image
smoothed with a Gaussian kernel of FWHM $6\farcs5$ after subtracting
the two background sources. Integrating all the emission inside a
circumference of $15\arcsec$ radius we find a total flux of
$5.0\pm1.2$ mJy (including the stellar emission and calibration
uncertainty), slightly lower but consistent within errors with the
previous data presented in \cite{Panic2013}.

%% , slightly lower
%% than the previous estimate but still consistent within the
%% uncertainty.

\section{ALMA Observations}
\label{sec:alma}

ALMA band 7 (0.86~mm) observations of 61~Vir were carried out on 2015
April, split into 4 scheduling blocks (one on April 9 and three on
April 22) as part of the project 2013.1.00359.S (PI: M.C. Wyatt). The
total number of antennas was 44, with baselines ranging from 15 to 349
m, with 5th and 95th percentiles equivalent to 29 and 228 m. This
allows us to recover angular scales of $0\farcs6$ up to $6\arcsec$ on
the sky.

The correlator was set up with three spectral windows to image the
continuum centered at 333.84, 335.78 and 347.74~GHz, each with 128
channels and a total bandwidth of 1.88~GHz; and a fourth one to search
for CO (v=0, 3-2) emission in the disc centered at 344.85~GHz, with 3840
channels, a channel width of 0.42~\kms\ (effective spectral
resolution of 0.82 \kms) and a total bandwidth of 2 GHz.

In all of the scheduling blocks J1337-1257 was used as Bandpass and
phase calibrator, with Titan as amplitude and flux
calibrator. Calibrations were applied using the pipeline provided by
ALMA. The total time on source excluding overheads was 178 min.

%% \begin{table*}
%%     \centering
%%     \caption{Summary of science observations.}
%% %    Remember to define the quantities, symbols and units used.}
%%     \label{sciobs}
%%     \begin{tabular}{lcccccccc} % four columns, alignment for each
%%         \hline
%% Date of observations & elevation [deg] & n$_\mathrm{ant}$ & t$_\mathrm{sci}$ [min] & $\%$ flagged & image rms [$\mu$Jy] & Flux calibrator & Bandpass calibrator & Phase Calibrator \\
%%         \hline
%%         2013 Dec 15 & 51-61 & 27 & 40 & 16 & 90 & Pallas   & J1058-0133 & J1215-1731 \\
%%         2013 Dec 15 & 70-79 & 27 & 40 & 40 & 90 & Ceres    & J1256-0547 & J1215-1731 \\
%%         2014 Dec 25 & 72-83 & 38 & 48 & 20 & 60 & 3c279    & J1256-0547 & J1245-1616 \\
%%         2014 Dec 26 & 77-83 & 40 & 48 & 18 & 40 & Ganymede & J1256-0547 & J1245-1616 \\
%%         2014 Dec 29 & 66-81 & 37 & 48 & 15 & 50 & Titan    & J1256-0547 & J1245-1616 \\
%%         2015 Jan 01 & 62-77 & 37 & 48 & 20 & 50 & Titan    & J1256-0547 & J1245-1616 \\
%%                 \hline
%%     \end{tabular}
%% \end{table*}

\subsection{Continuum emission}
\label{sec:continuum}

\begin{figure}
    
  \includegraphics[trim=0.0cm 0.5cm 0.0cm 1.0cm, clip=true,
    width=1.0\columnwidth]{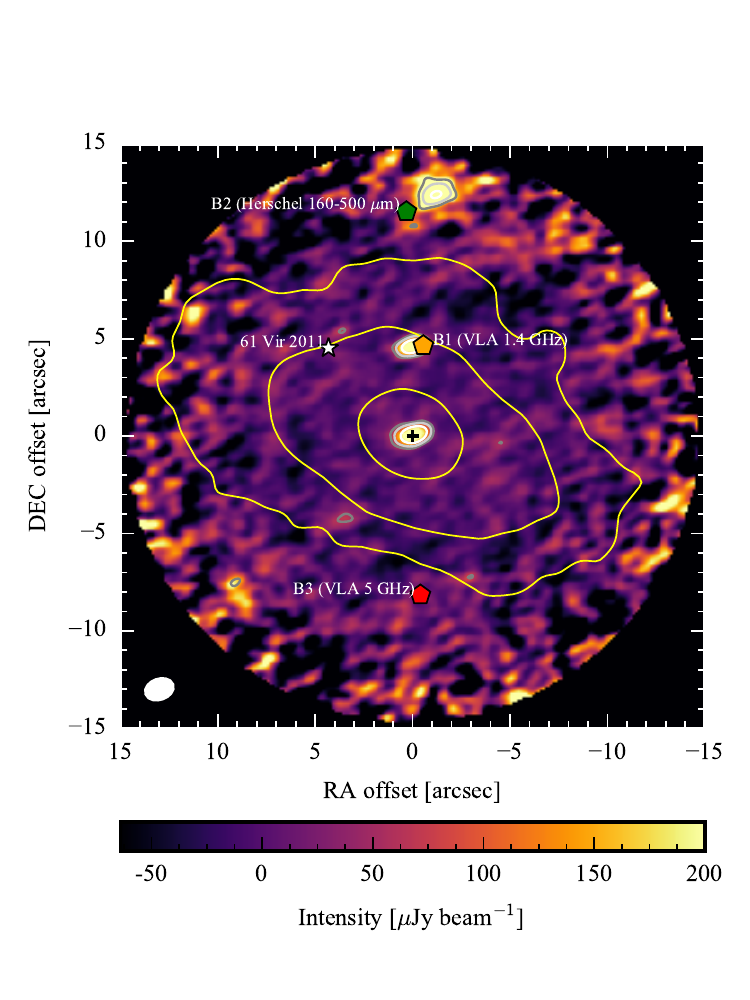}

  \caption{ALMA band 7 (0.86~mm) continuum image of 61~Vir with
    natural weights and corrected by the primary beam response
    (FWHM$\sim17\arcsec$). The beam size is $1\farcs1\times0\farcs7$
    and is represented with a white ellipse at the bottom left corner
    of the image. The grey and white contours represent emission above
    3, 5 and 10 times the local noise level. Yellow contours from the
    Herschel 70 $\mu$m image at arbitrary levels are overlayed
    correcting by the stellar proper motion. The x- and y-axes
    indicate the offset from the stellar position in R.A. and decl. in
    arcsec, i.e. north is up and east is left. The stellar position is
    marked with a black ``+'' and the position of background sources
    previously detected are represented with pentagons. The black
    masked region indicates a primary beam response below 10\%}
  \label{fig:continuum}
  
\end{figure}

To study the continuum emission, we use the four spectral windows to
reach the highest sensitivity as no CO emission is present in the data
(this is discussed below). Figure \ref{fig:continuum} shows the
continuum image using the task \textsc{CLEAN} in \textsc{CASA} 4.4
\citep{McMullin2007} with natural weights and correcting for the
primary beam --- Note that the noise increases towards the edges of
the image as the primary beam sensitivity decreases. At the center of
the image we achieve a rms noise level of 16~$\mu$Jy~beam$^{-1}$,
which increases to 32~$\mu$Jy~beam$^{-1}$ at $7\farcs5$. The beam size
is $1\farcs1\times0\farcs7$ with a position angle (PA) of
$-70^{\circ}$.  In the image three compact sources are detected:
61~Vir's stellar emission at the center with a total flux of
$374\pm16\ \mu$Jy, which is $2.4\sigma$ higher than the
$320\pm16$~$\mu$Jy predicted photospheric emission assuming a spectral
index of -2, thus, it could be due to chromospheric emission at this
wavelength \citep[e.g.,][]{Loukitcheva2004, Fontenla2007}; and two
other sources to the north of the star with offsets of $4\farcs5$ and
$12\farcs5$, and peak fluxes of $360\pm20\ \mu$Jy and
$850\pm70\ \mu$Jy, respectively. The latter is resolved with a total
flux of $2.2\pm0.3$ mJy within a $2\arcsec$ radius
circumference. These two sources are almost certainly the background
galaxies previously reported in \cite{Wyatt2012} and their position is
overlayed with pentagon markers and labelled as B1 and B2. We also
overlay the position of a third background source (B3) detected at 5
GHz with the VLA and not present in the ALMA data. The latter is the
southern component of a double-lobed structure with the northern
component outside the ALMA primary beam. At 1.4 GHz, B1 was marginally
resolved and found to be extended in the north-south direction with a
fitted FWHM of $33\arcsec$, therefore, consistent as being the two
lobes resolved at 5 GHz.

\begin{figure}
    
  \includegraphics[trim=0.0cm 0.0cm 0.0cm 0.0cm, clip=true,
    width=1.0\columnwidth]{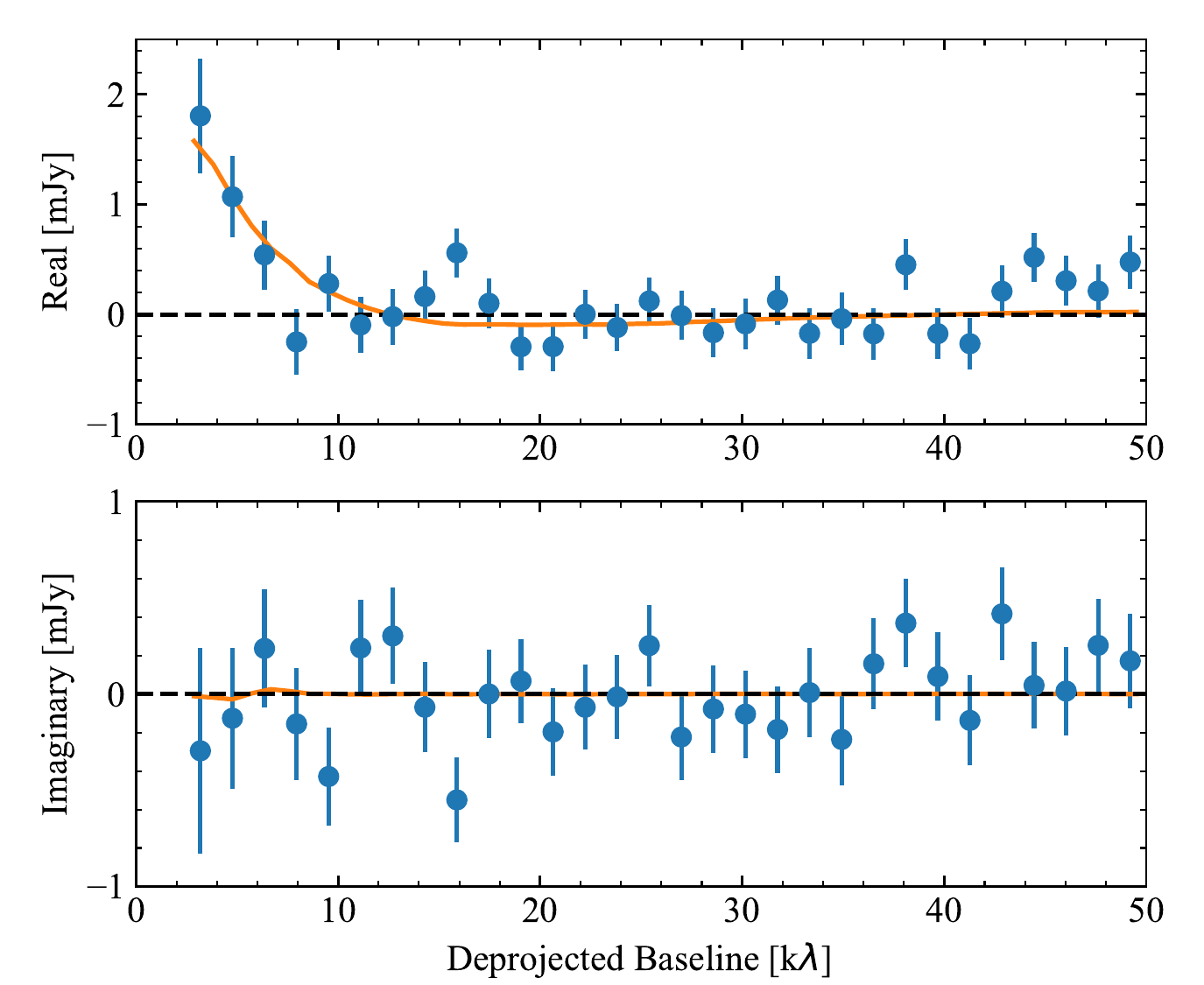}

  \caption{Deprojected visibility profile of the ALMA band 7 (0.86~mm)
    continuum after subtracting the emission from the three compact
    sources. The blue points represent averaged and binned
    visibilities with 1$\sigma$ errorbars. Overlayed is a best fit
    disc model (orange line).}
  \label{fig:vis}
\end{figure}

Although there is no disc emission above $3\sigma$ in the ALMA CLEAN
image, significant signal is present in the real component of
  the visibilities after subtracting the three compact sources. By
  de-projecting the observed visibilities assuming a disc PA and
  inclination of $65\degr$ and $77\degr$, respectively
  \citep[consistent with the Herschel observations,][]{Wyatt2012}, we
  recover disc emission in the short baselines ($\lesssim10$
  k$\lambda$, see Figure \ref{fig:vis}), corresponding to extended
  emission ($\gtrsim20\arcsec$ or 150 AU). We also overlay the model
  visibilities of a disc with a flux of 4 mJy and extending from 30 to
  140 AU, consistent with the data (see Sec. \ref{sec:model}). The
  imaginary part of the visibilities is consistent with pure noise
  around zero, which is expected for an axisymmetric centered disc.

\begin{figure}

  \begin{subfigure}%[]{0.5\columnwidth}

    \includegraphics[trim=0.0cm 0.0cm 1.0cm 1.0cm, clip=true,
    width=1.0\columnwidth]{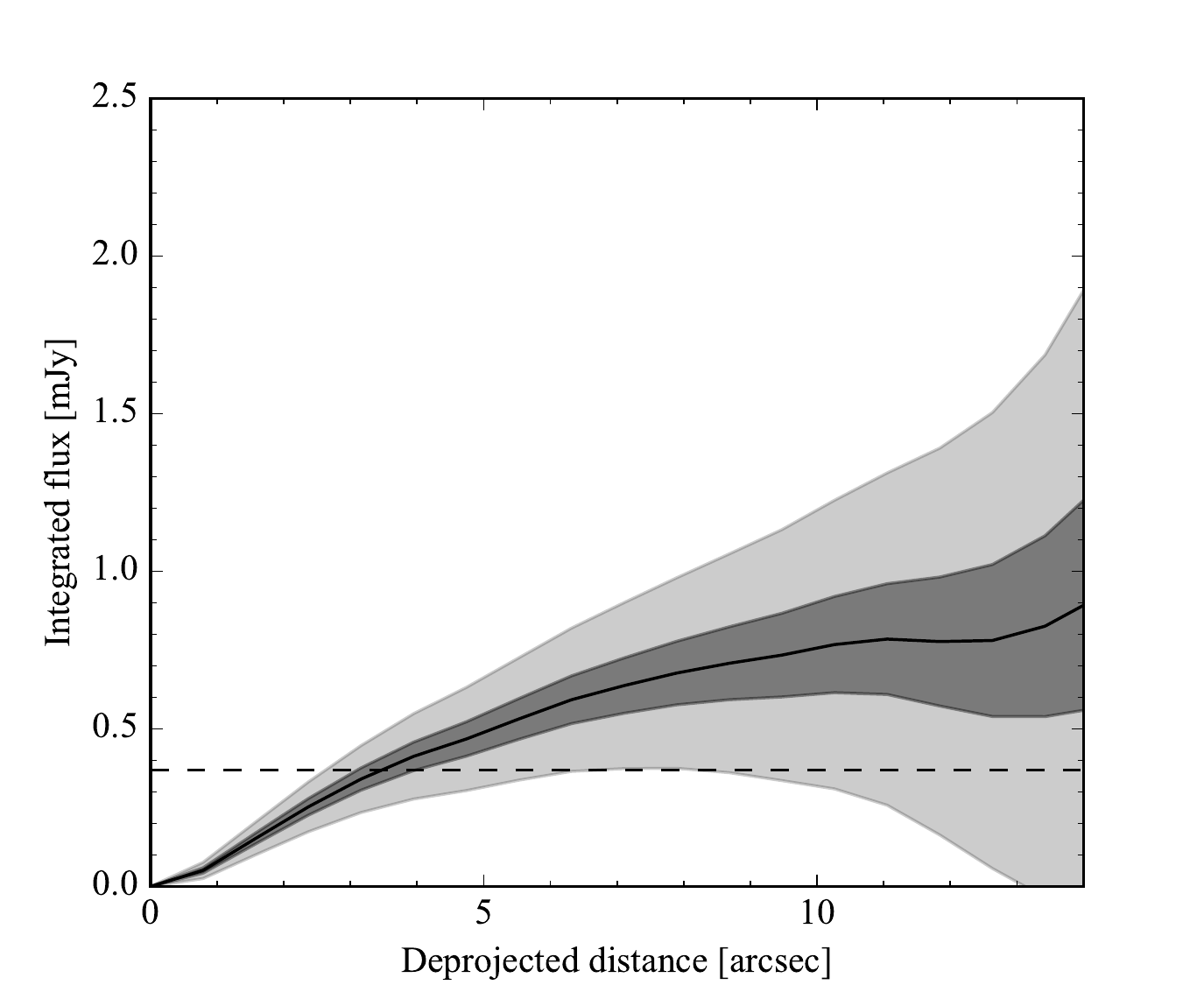}

  \end{subfigure}

  \begin{subfigure}%[]{1.0\columnwidth}

    \includegraphics[trim=0.0cm 0.0cm 1.0cm 1.0cm, clip=true,
    width=1.0\columnwidth]{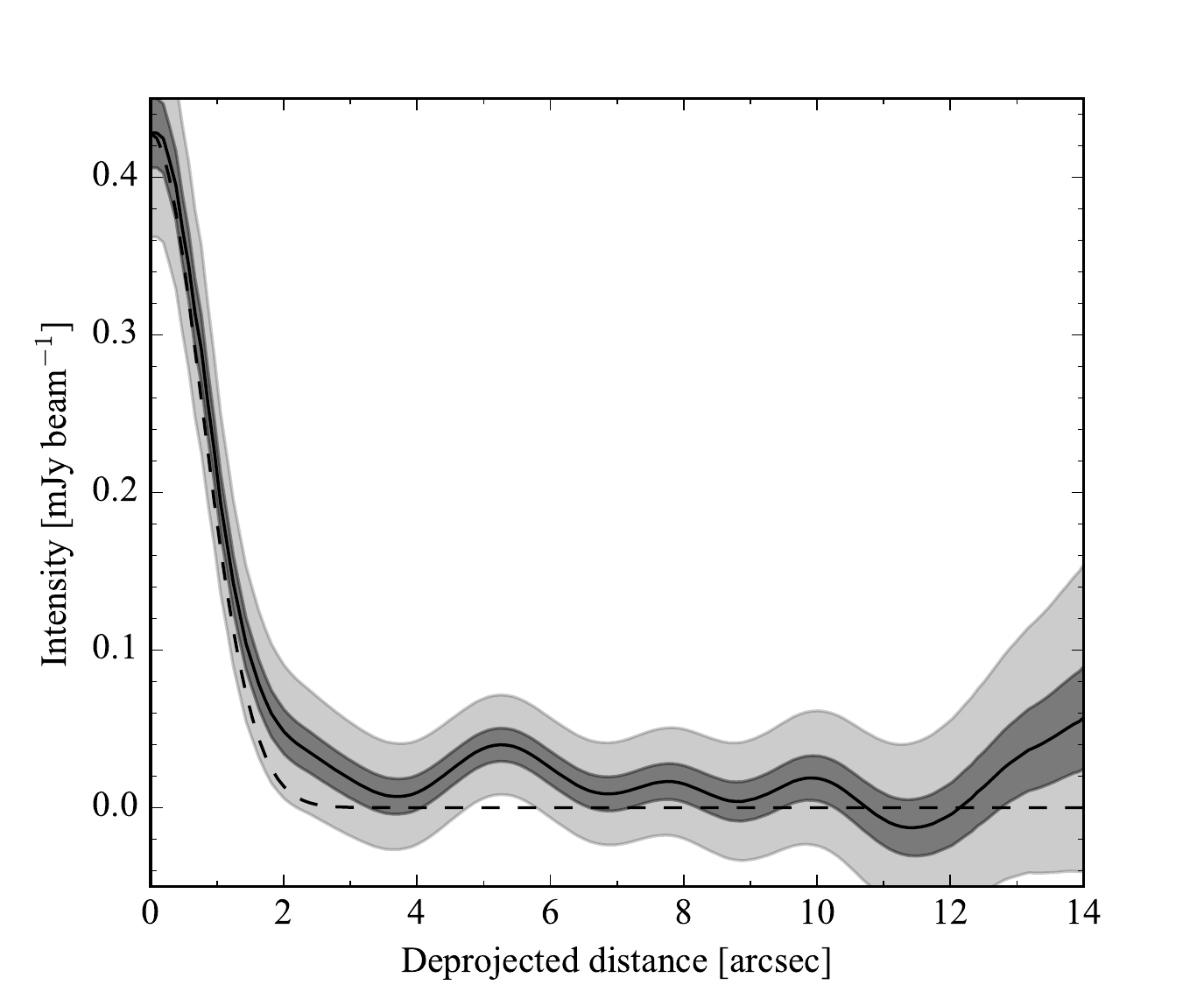}

  \end{subfigure}

  \caption{Top: Integrated flux vs semi-major axis of elliptic regions
    over which the flux is integrated. The dashed line represents the
    stellar flux. Bottom: Average intensity radial profile computed
    in wedges of $\pm30\degr$ along the major axis of the disc, using the
    reconstructed CLEAN image tapered with the Fourier transform of a
    Gaussian of FWHM of $1\farcs5$. The dashed line represents the
    PSF. The grey shaded areas in both panels represent 68\% and
    99.7\% confidence regions.}
    \label{fig:Ir}
\end{figure}

We can also recover the disc emission in the image space by
integrating the flux inside ellipses of different semi-major axes
(with the same PA and aspect ratio or inclination as the disc resolved
by Herschel). In this integration we also exclude a $30\degr$ wide
wedge in the direction of B1. The resulting radial profile is
presented in the top panel of Figure \ref{fig:Ir}. Within $10\arcsec$
the total disc and stellar emission is only $0.8\pm0.2$~mJy, 2.2
$\sigma$ lower than the derived flux from SCUBA2. If we subtract the
stellar emission, the disc is marginally detected at $2.2\sigma$ with
a total flux of $0.43\pm0.2$ mJy. The lower ALMA disc flux could be
produced by spatial filtering in the ALMA data due to a lack of short
baselines, as the maximum recoverable scale is $6\arcsec$ given the
range of baselines in the data. This is illustrated in Figure
\ref{fig:vis} and demonstrated in Sec. \ref{sec:model}, where we fit
and simulate the observed visibilities and the SCUBA2 image using a
parametric disc model that we use to constrain the disc flux and disc
surface density.

%% Another possibility is that the SCUBA2 data was contaminated by two
%% background galaxies, therefore in Sec. \ref{sec:scuba} we present
%% new 0.88 mm photometry subtracting the background sources inside
%% the SCUBA2 primary beam.

We search for any spatially resolved disc emission by azimuthally
averaging a CLEAN image of the ALMA data spatially smoothing the
emission tapering the visibilities with the Fourier transform of a
Gaussian of FWHM of $1\farcs5$. This process degrades the CLEAN beam
to a size of $1\farcs8\times1\farcs5$ increasing the Signal-to-noise
ratio (S/N) for extended emission. The azimuthal averaging method also
takes into account the disc inclination and PA and is done in wedges
of $\pm30\degr$ along the major axis of the disc. At each radius, the
uncertainty is computed based on the uncertainty on each pixel and the
number of independent measurements, estimated to be equal to the
length of the arc over which we are averaging, divided by the beam's
semi-major axis. The azimuthally averaged intensity is presented in
the bottom panel of Figure \ref{fig:Ir}. This shows a marginal disc
detection of $0.04\pm0.01$ mJy~beam$^{-1}$ at $5\farcs5\pm0\farcs9$
(where the positional uncertainty is roughly estimated as $\sim$beam
semi-major axis/$\sqrt{S/N}$), equivalent to $47\pm8$ AU, consistent
with the inner disc radius constrained to be between $30-40$ AU,
depending on the disc model assumed to fit the Herschel observations
\citep{Wyatt2012}. Moreover, positive emission, but not significantly
above zero apart from the peak at $5\farcs5$, is present from the
stellar position to a distance of $11\arcsec$. This is consistent with
the positive total flux described before, in other words, with the
$2.2\sigma$ detection integrated over all radii.

%% \subsection{CO (3-2)}
%% \label{sec:co}

\subsection{CO}

Although CO gas of secondary origin has been found in a few young
bright debris discs, probably released in collisions of icy solids
\citep[e.g., 49 Ceti, $\beta$~Pic, HD~131835, HD~181327 and
  Fomalhaut,][]{Zuckerman1995, Dent2014, Moor2015gas, Marino2016,
  Matra2017betapic, Matra2017Fomalhaut} no CO (v=0, J=3-2) emission
was detected in 61 Vir ALMA data. Integrating the continuum subtracted
channel maps from 30 to 100 AU and radial velocities (RV) in the range
$\pm5.1$ \kms with respect to the stellar RV (expected Doppler shift
due to Keplerian rotation at 30 AU), we derive an integrated noise
level of 27 mJy~\kms. We can use this to place a $3\sigma$ upper limit
to any CO present in the disc. As shown by \cite{Matra2015} non-local
thermodynamic equilibrium (non-LTE) effects can be significant in the
low density environments of debris discs; therefore, it is necessary
to consider the effect of different gas kinetic temperatures and
collisional partner densities --- assumed to be electrons released
from carbon ionization after the CO gas is photodissociated \citep[as
  predicted by thermodynamic models, e.g.,][]{Kral2016}. Using the
tools developed by \cite{Matra2015} we derive a CO gas mass upper
limit of $1.4\times10^{-6}$~\Mearth \ using the $3\sigma$ upper limit
on the CO flux, the assumed disc extent (30-100~AU) and a carbon
ionization fraction of 0.5 and a C/CO abundance of 100 \citep[assumed
  to be equal to those in $\beta$~Pic,][]{Cataldi2014, Roberge2000},
which fixes the ratio between electron and CO gas number densities in
the disc.

Given the short photodissociation timescale of 120~yr, together with
the low dust optical depth, and thus, low collisional rates of solids
in the disc, we do not expect to detect CO gas being released in
collisions of icy planetesimals in this system. For example, if we
assume that planetesimals in the disc have a CO mass fraction of 16\%,
near the maximum fraction that has been observed in Solar System
comets \citep[0.3-16\%,][]{Mumma2011} and similar to other systems
with detected exocometary gas \citep{Matra2017Fomalhautinprep}, we
expect only $\lesssim10^{-9}$~\Mearth \ of CO gas in the disc coming
out from collisions. Greater amounts of CO gas trapped in ices could
come out from icy planetesimals closer in if these are scattered into
highly eccentric orbits that can cross the H$_2$O or CO$_2$ snow lines
within 10 AU, as suggested by recent ALMA observations of $\eta$~Corvi
\citep{Marino2016}, but this is not detected and no evidence of such
scattering has been found so far for 61 Vir.

\section{Disc modelling}
\label{sec:model}

In order to place better constraints on the total disc flux, disc
size, inclination and position angle, we fit a parametric disc model
to the SCUBA2 image and ALMA visibility data simultaneously. The model
consists of a central star surrounded by a dusty disc and two
background point sources (B1 and B2) at the position of the maxima in
the ALMA image. The fluxes of the star, B1 and B2 are held fixed at
their observed values of 0.37, 0.36 and 0.85 mJy, respectively. Note
that the B2 is apparently resolved and could be modelled with an
extended component, but this has no effect on the fitted parameters
and best fit models.

The dusty disc is assumed to be composed of grains formed by
astrosilicates \citep{Draine2003}, amorphous carbon
\citep{LiGreenberg1998}, and water ice \citep{LiGreenberg1998}, with
mass fractions of 70\%, 15\% and 15\%, respectively. We mix the
optical constants using the Bruggeman rule \citep{BohrenHuffman1983}
and mass-weighted opacities are computed using the Mie theory code of
\cite{BohrenHuffman1983}, assuming a Dohnanyi-like size distribution
with a power law index of -3.5 \citep{Dohnanyi1969}, and minimum grain
size of 1 $\mu$m, roughly the blow-out size, and a maximum size of 1
cm. We expect larger grains to be present, but we can neglect their
thermal emission at this wavelength. The central star is modelled
using a stellar template spectrum with a effective temperature of 5500
K\footnote{http://www.stsci.edu/hst/observatory/crds/k93models.html}
\citep{Kurucz1979} and a radius of 1.1 $R_\odot$ to fit the stellar
emission at 0.86~mm. Then, the dust equilibrium temperature at
different radii is computed using
RADMC-3D\footnote{http://www.ita.uni-heidelberg.de/$\sim$dullemond/software/radmc-3d/}
\citep{RADMC3D0.40}. The disc surface density varies with radius and
is parametrized with a power law function as $r^{\alpha}$ from a
minimum radius of 30 AU, extending to $R_{\max}$, which is a free
parameter as well as $\alpha$ and the total disc flux,
$F_\mathrm{disc}$. We maintain $R_{\min}$ fixed at 30 AU \citep[best
  fit value for a model with a sharp inner edge when fitting the
  Herschel observations and SED,][]{Wyatt2012}. The vertical mass
distribution is assumed to be Gaussian with a standard deviation or
scale height H that scales linearly with radius as H$=0.1r$.
Synthetic images at 0.86~mm are then produced using RADMC-3D with an
inclination, $i$, and PA that are also left as free parameters. In
total there are 5 free parameters that we vary to fit the
observations.

Model visibilities are computed at the same uv points as the ALMA
observations \citep[e.g., ][]{Marino2015mwc, Marino2016,
  Marino2017}. To speed up the simulation of model visibilities, we
average the ALMA data with a time and frequency bin of 90~s and 1.88
GHz, respectively. This averaging is small enough both in time and
frequency to ensure that the time and frequency smearing are smaller
than $0\farcs1$ ($\ll$ synthesised beam). We simulate the SCUBA2
observation by convolving the model image with a two dimensional
Gaussian with a FWHM of $13\arcsec$.

To find the best fit we use a Bayesian approach, sampling the
parameter space using the python module \textsc{emcee}, which
implements Goodman \& Weare's Affine Invariant MCMC Ensemble sampler
\citep{GoodmanWeare2010, emcee}. The posterior distribution is defined
as the product between the likelihood function and our prior
distributions. The first is defined as $\exp(-\chi^{2}/2)$, with
$\chi^2=\chi^2_{\rm ALMA}+\chi^2_{\rm SCUBA2}$, with
\begin{equation}
  \chi_{\rm ALMA}^2=\sum_{i} \frac{||V_\mathrm{data, i}-V_\mathrm{model, i}||^{2}}{\delta V_\mathrm{data, i}^2},
\end{equation}
where the sum goes over the uv points of the previously averaged
visibilities, $V_\mathrm{data, i}$. The estimated error $\delta
V_\mathrm{data,i}$ is calculated based on the intrinsic dispersion of
the visibilities over one scan with the task \textsc{statwt} from
\textsc{casa} 4.7. On the other hand, $\chi^2_{\rm SCUBA2}$ is defined
as the squared sum over every pixel of the difference between the
SCUBA2 and model image (convolved with the $13\arcsec$ beam), divided
by the pixel rms. The pixel rms is empirically estimated by measuring
the dispersion on the unsmoothed SCUBA image that has uncorrelated
pixel noise.

The prior probabilities of the parameters are assumed to be
uniform. We restrict $R_{\max}$ to be between 30 and 250 AU, $\alpha$
from $-5$ to 5, $F_\mathrm{disc}$>0, PA from $0^{\circ}$ to
$90^{\circ}$ and $i$ from $45^{\circ}$ to $90^{\circ}$ (priors based
on the previous Herschel observations).

\begin{figure}
  \includegraphics[trim=0.0cm 0.0cm 1.0cm 0.0cm, clip=true,
    width=1.0\columnwidth]{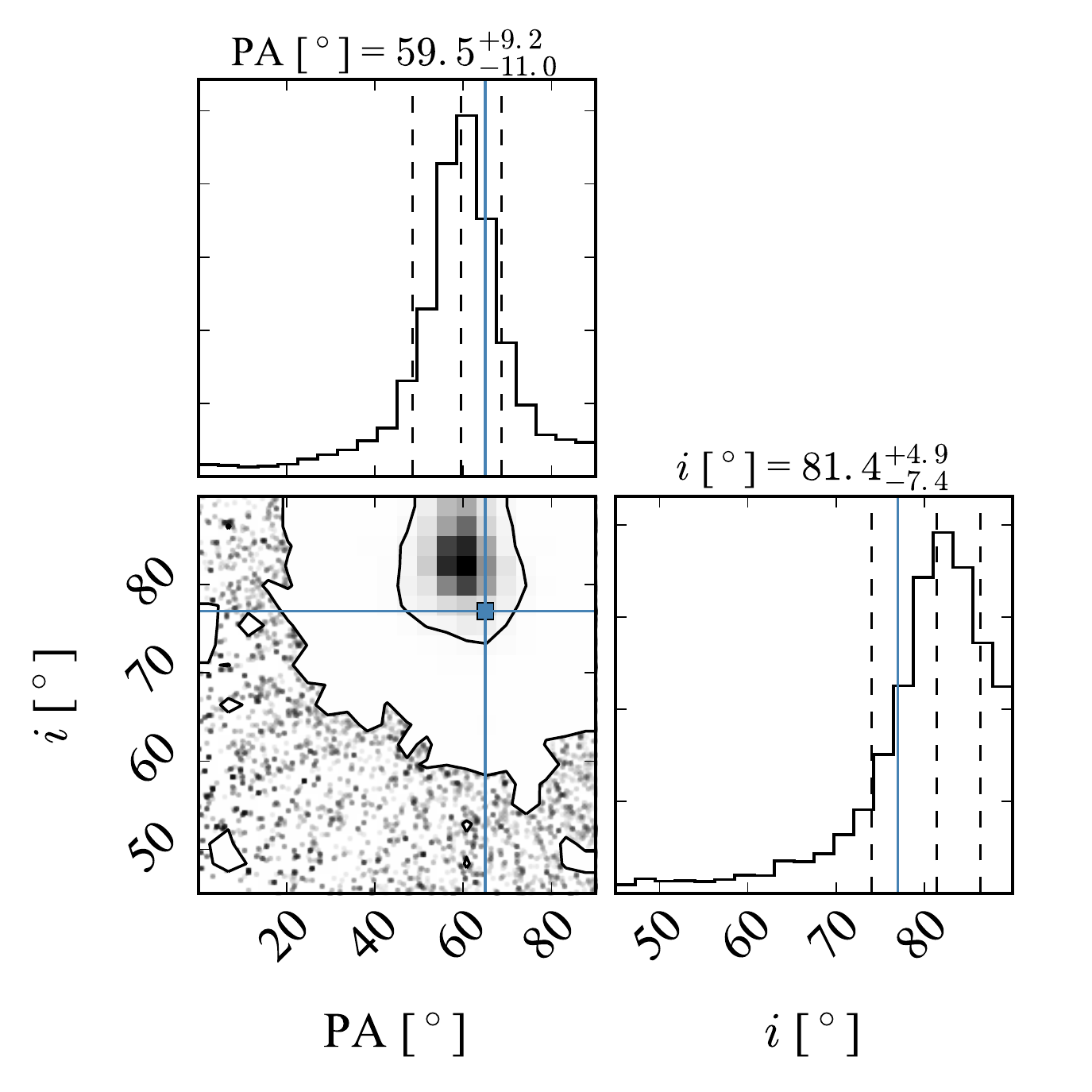}
  \caption{Posterior distributions of PA and $i$ when fitting the ALMA
    data only.  The vertical dashed lines represent the 16th, 50th and
    84th percentiles. Contours correspond to 68\% and 95\% confidence
    regions. The blue lines represent the previous estimates of $i$
    and PA from Herschel observations. This plot was generated using
    the python module \textsc{corner} \citep{cornerplot}.  }
    \label{fig:painc}
\end{figure}

To demonstrate that there is disc emission in the ALMA data that can
be better constrained by adding the SCUBA2 image to the fitting
process, in Figure \ref{fig:painc} we present the marginalised
distributions of $i$ and PA when fitting only the ALMA data and
constraining $R_{\max}$ to values below 140 AU as any disc emission
beyond that would lie outside the ALMA primary beam. Even though disc
emission above 3$\sigma$ is not present in the reconstructed ALMA
image (see Figure \ref{fig:continuum}), but only when integrating the
emission, we find that the disc orientation can still be constrained
and matches with the previous estimates from Herschel observations
(blue lines).

\begin{figure}
  \includegraphics[trim=0.0cm 0.0cm 0.0cm 0.0cm, clip=true,
    width=1.0\columnwidth]{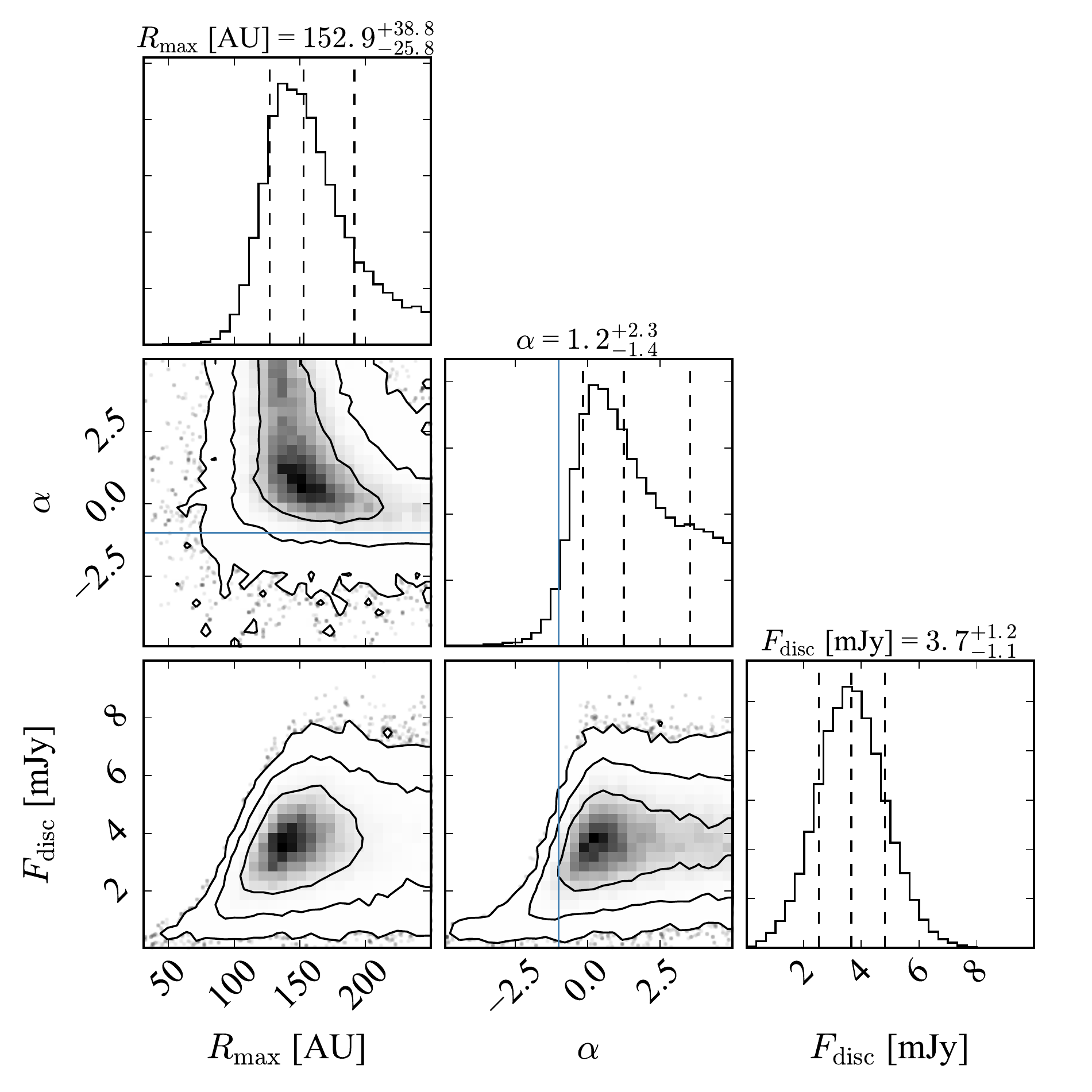}
  \caption{Posterior distribution of $R_{\max}$, $\alpha$ and
    $F_\mathrm{disc}$. The vertical dashed lines represent the 16th,
    50th and 84th percentiles. Contours correspond to 68\%, 95\% and
    99.7\% confidence regions. The blue lines represent the previous
    estimate of $\alpha$ from Herschel observations. This plot was
    generated using the python module \textsc{corner}
    \citep{cornerplot}.  }
    \label{fig:mcmc}
\end{figure}

Figure \ref{fig:mcmc} presents the marginalised distributions of
$R_{\max}$, $\alpha$ and $F_\mathrm{disc}$, when ALMA visibilities and
the SCUBA2 image are combined in the analysis. The disc orientation is
better constrained, with PA$=59\pm5$ and $i=82\pm4^{\circ}$,
consistent with the Herschel observations (PA$=65^{\circ}$ and
$i=77^{\circ}$), and within the limits obtained from fitting the ALMA
data alone (see Figure \ref{fig:painc}). Regarding the disc structure,
we find that $\alpha$ peaks at zero on its marginalised posterior
distribution and is constrained between -0.2 and 3.5 (68\%
confidence), but still consistent within the 95\% confidence region
with the value of -1 (see Figure \ref{fig:mcmc}) found by fitting the
Herschel observations, which was also poorly constrained
\citep{Wyatt2012}. If we restrict $i$ between $70-80^{\circ}$ (using
the prior information from Herschel images), we can improve our
constraints on the slope, finding
$\alpha=0.1^{+1.1}_{-0.8}$. Therefore, we conclude that the surface
density distribution is not very centrally concentrated.

For example, we can discard a scattered disc that has an initial
characteristic surface density proportional to $r^{-3.5}$
\citep[e.g.,][]{Duncan1997}. The collisional evolution of such an scattered
disc has been studied analytically by \cite{Wyatt2010}. We find that
for 61 Vir parameters, i.e. assuming $t=4.6 Gyr$, $\alpha=2.5$,
pericentre at 30~AU and $M_{\rm disc}\sim10^{-2}-10^2$~M$_\oplus$, the
resulting surface density should be significantly peaked at 30 AU
(pericentre) and decrease steeply with radii, inconsistent with our
observations (see their Figure 5). Although the analytic model used by
\cite{Wyatt2010} could overestimate the surface density of dust at low
radii as it is the case for low eccentricities.

On the other hand, a flat distribution could be expected in the
context of an extended disc with a wide range of semi-major axes and
small eccentricities, collisionally evolved after being stirred
\citep[e.g., ][see Sec. \ref{sec:collmodel}]{Schuppler2016,
  Geiler2017}. We also find that $R_{\max}$ is peaked at $\sim150$~AU,
consistent with the maximum radius of at least 100 AU derived with
Herschel. However, if $\alpha<0.5$ then the maximum radius is not well
constrained as the surface brightness decreases with radius
($B(r)\propto r^{\alpha-0.5}$).  $F_\mathrm{disc}$ peaks above zero
($3.4\sigma$), and is constrained to be $3.7^{+1.2}_{-1.1}$ mJy;
however, this is highly dependent on $R_{\max}$ and $\alpha$. For
example, if $\alpha\sim -1$ then $F_\mathrm{disc}<4$ mJy (95\%
confidence).

We also try to vary $R_{\min}$ and leave it as a free parameter, but
we find that it is not well constrained in these observations. The
posterior marginalised distribution of $R_{\min}$ is close to flat with
a peak at the inner boundary set to 5 AU. With a smaller $R_{\min}$
the disc surface brightness decreases which fits best the ALMA
visibilities, while conserving the total flux to fit the SCUBA
observations. Therefore, we decide to leave $R_{\min}$ fixed based on
the previous Herschel and SED information that are inconsistent with
$R_{\min}\ll30$~AU.

%% Moreover, if we restrict $i$ between $60-80^{\circ}$
%% we find $\alpha=0.1^{+1.1}_{-0.8}$, closer to the derived surface
%% density using Herschel observations ($\alpha\sim-1$). A flat
%% distribution could be expected in the context of an extended disc
%% collisionally evolved after being stirred \citep[e.g., ][see
%%   Sec. \ref{sec:collmodel}]{Schuppler2016}.

\begin{table}
  \centering
  \caption{Best fit values of the ALMA and SCUBA2 data
    combined. Median $\pm$ uncertainty based on the 16th and 84th
    percentile of the marginalised distributions.}
  \label{table:gaussian}
  \begin{tabular}{cc} %cc} % 7 columns, r,dr,h,pa,inc,raoff,decoff 
    \hline
    \hline
    Parameter & Best fit value \\
    \hline
    %% $R_\star$ [$R_\odot$] & $1.46\pm0.08$ \\
    %% $M_{d}$ [M$_{\oplus}$] & $0.014\pm0.001$  \\
    $R_{\max}$ [AU] & $153^{+39}_{-26}$ \\
    $\alpha$  &  $1.2^{+2.3}_{-1.4}$  \\
    $F_\mathrm{disc}$ [mJy] &  $3.7^{+1.2}_{-1.1}$  \\
    PA [$^{\circ}$] &  $59\pm5$ \\
    $i$ [$^{\circ}$] & $82\pm4$ \\
    %% RA offset [$\arcsec$] & $-0.03\pm0.08$  \\
    %% Dec offset [$\arcsec$] &  $0.00\pm0.07$ \\
    \hline
  \end{tabular}
\end{table}

%% \begin{figure*}
%%   \includegraphics[trim=0.0cm 0.0cm 0.0cm 0.0cm, clip=true,
%%     width=0.7\textwidth]{corner_all.pdf}
%%   \caption{Posterior distribution of $R_{\max}$, $\alpha$,
%%     $F_\mathrm{disc}$, PA and $i$.  The vertical dashed lines
%%     represent the 16th, 50th and 84th percentiles. Contours correspond
%%     to 68\% and 95\% confidence regions. The blue lines represent the
%%     previous estimates of $\alpha$, $i$ and PA from Herschel
%%     observations. This plot was generated using the python module
%%     \textsc{corner} \citep{cornerplot}.  }
%%     \label{fig:mcmc}
%% \end{figure*}

\begin{figure*}
  \includegraphics[trim=1.0cm 0.5cm 1.0cm 0.5cm, clip=true,
    width=1.0\textwidth]{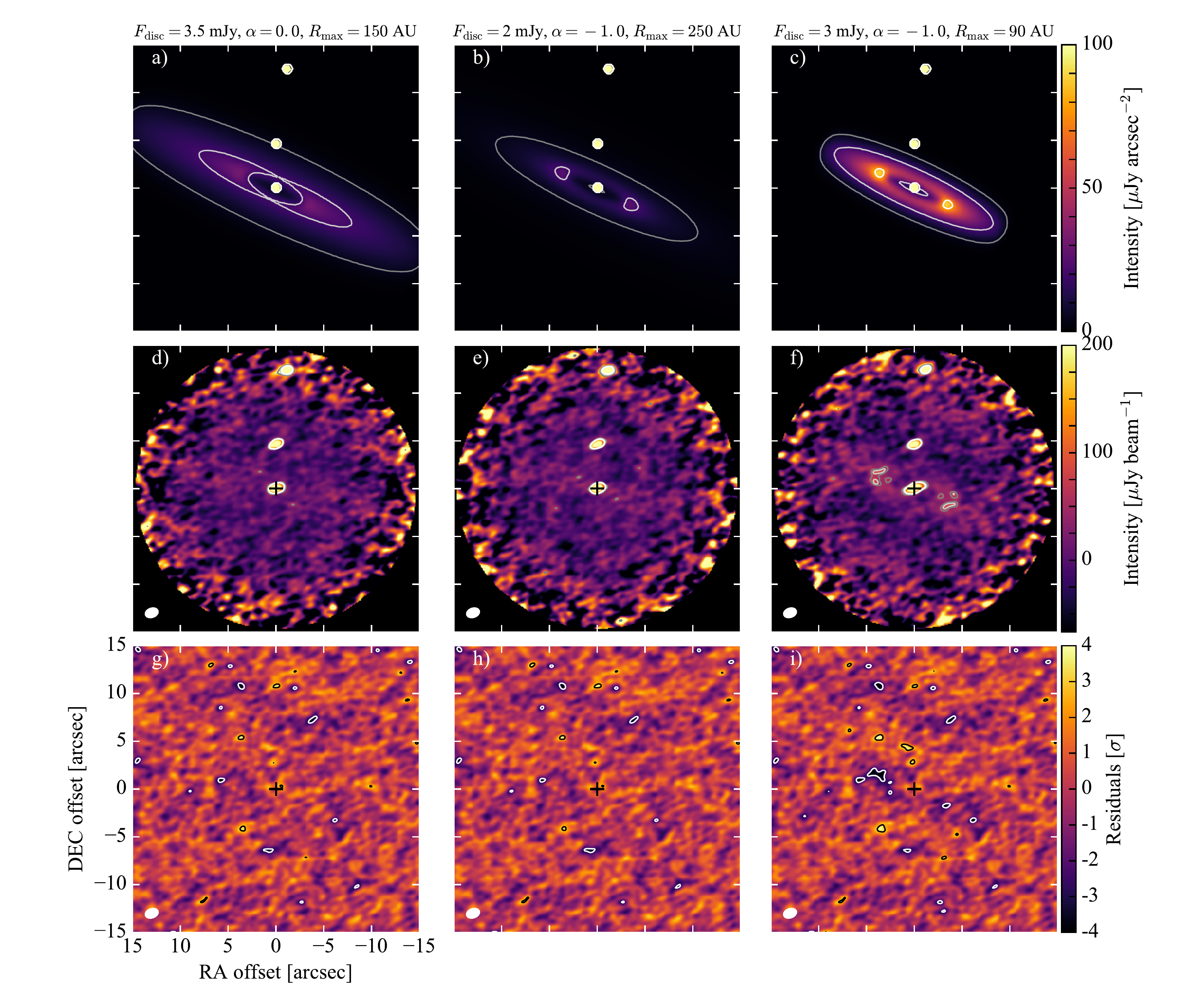}
  \caption{Simulated model images and residuals at 0.86~mm. First
    column: $F_\mathrm{disc}=$3 mJy, $\alpha=0$, PA$=65^{\circ}$ and
    $i=77^{\circ}$. Second column: $F_\mathrm{disc}=$3.5 mJy,
    $\alpha=-1$, PA$=65^{\circ}$ and $i=77^{\circ}$. Third column:
    $F_\mathrm{disc}=$6 mJy, $\alpha=-1$, PA$=65^{\circ}$ and
    $i=77^{\circ}$. First row: synthetic images of the disc. Contours
    represent 5, 20 and 80 $\mu$Jy~arcsec$^{-2}$. Second row: primary
    beam corrected simulated ALMA CLEAN images using the same
    uv-sampling and adding Gaussian noise to the visibilities,
    according to their variance in the observations. Contours
    represent 3, 4 and 5 times the local noise level. Third row: Dirty
    map of the ALMA residuals after subtracting the model visibilities
    from the ALMA observations. The noise level on the residuals is
    uniform and equal to 16~$\mu$Jy~beam$^{-1}$ as they are not
    corrected by the primary beam. The black and white contours
    represent $\pm3\sigma$. The beam size is represented by a white
    ellipse in the bottom left corner. The x- and y-axes indicate the
    offset from the stellar position in R.A. and decl. in arcsec,
    i.e. north is up and east is left. The stellar position is marked
    with a black ``+''.}
    \label{fig:models}
\end{figure*}

In Figure \ref{fig:models} we compare simulated observations of
different models and their residuals when subtracted from the real
observations. The first column shows the best fit model from the
posterior distribution presented above with a total flux of 3.5 mJy,
$\alpha=0$, $R_{\max}=150$~AU, PA$=65^{\circ}$ and $i=77^{\circ}$,
which has a reduced chi-squared $\chi^2_\mathrm{red}$=1.0028838
($N\sim6\times10^{6}$). The second column shows a model with
$\alpha=-1.0$, $R_{\max}=250$ AU and $F_\mathrm{disc}=2$ mJy, i.e. the
most likely disc flux for this $\alpha$. This model is still
consistent with having no disc emission above $3\sigma$ in the
reconstructed image and has $\chi^2_\mathrm{red}$=1.0028840
($1.4\sigma$ difference with the first model). The third column
corresponds to a model similar to the second, but with a less extended
disc with $R_{\max}=90$~AU and $F_\mathrm{disc}=3$ mJy, increasing the
surface brightness of the disc to levels above $3\sigma$ in the
simulated observation (Figure \ref{fig:models}f), which translates to
significant negative residuals (Figure \ref{fig:models}i) and
$\chi^2_\mathrm{red}$=1.0028897 ($35\sigma$ difference with the
previous model). We also find that the image reconstruction suffers
from flux loss due to an insufficient number of short baselines and
the size of the primary beam ($17\arcsec$). For the models in the 1st,
2nd and 3rd columns, we recover integrated fluxes of 0.6, 1.0 and
$2.3\pm0.2$ mJy, respectively. From the best fit values of the
parameters (i.e. $R_{\max}\gtrsim150$ AU), the SCUBA2 measured flux
($5.0\pm1.2$~mJy) and the simulated observations which show that a
compact disc would be detectable, we conclude that the disc of
planetesimals must be broad and not concentrated in a single or a few
narrow rings, which could not have been resolved by Herschel (model 3
in Sec. \ref{sec:intro}).

\section{Steady state collisionally evolved disc model}
\label{sec:collmodel}

It is generally assumed in debris discs that the surface density of
millimetre-sized grains can be simply scaled to derive the
distribution of the total solid mass in discs. This is true under the
assumption that the size distribution from big to small bodies remains
fixed. However, using detailed numerical simulations with the Analysis
of Collisional Evolution (\textsc{ACE}) code \citep{Krivov2006},
\cite{Schuppler2016} recently showed that the radial profile of the
vertical optical depth can deviate considerably from the distribution
of planetesimals, when considering this more realistic model of how
the grain size distribution evolves at different radii. For example,
when assuming a maximum planetesimal size of 100 and 200 km in
diameter, they found that the optical depth (dominated by the smallest
grains in the disc) stays roughly constant as a function of radius
between 10 and 100 AU, even though the total surface density decreases
with radius. This effect is not due to radiation pressure affecting
small dust grains, but due to the evolution of the size distribution
at different radii. Specifically, the difference arises when the
largest planetesimals in the disc (that dominate the disc mass) are
not collisionally evolved, but the smallest grains are already in
collisional equilibrium. Using a three phase analytic model for the
size distribution, \cite{Geiler2017} confirmed this effect and
explored how it changes depending on the primordial conditions of the
disc.

This implies that even if we assume that the primordial distribution
of solids in a debris disc is close to a standard Minimum Mass Solar
Nebula (MMSN) with a radial distribution with an exponent of -1.5
after the protoplanetary disc disperses \citep{Widenschilling1977mmsn,
  Hayashi1981}, or any model for the initial surface density profile
of an accreting protoplanetary disc \citep[e.g.,][]{Kuchner2004,
  Raymond2005, Chiang2013}, the radial distribution of dust grains
with lifetimes shorter than the age of the system could have a
significantly different radial dependence. Therefore, the surface
density exponent for millimetre grains derived in Sec. \ref{sec:model}
cannot be simply extrapolated to the total surface density of solids
in 61~Vir.

Here, we aim to study the expected surface density of millimetre
grains in a broad debris disc undergoing collisional evolution, and
how that depends on the choice of maximum planetesimal size. We do
this by using a simple numerical prescription that simulates the size
distribution using size bins and assuming that the size distribution
is in quasi steady state. This means that the mass loss rate due to
catastrophic collisions in each size bin is balanced by the input from
fragmentation of larger bodies in destructive collisions, which inputs
mass into the bin. The maximum size in collisional equilibrium, $D_c$,
corresponds to the one having a collisional lifetime equal to the age
of the system. This method is described in detail in \cite{Wyatt2011}
(see Sec. 2.4.2, 2.5 and 2.9 therein) and can reproduce the morphology
(slope and wiggles) seen in more detailed numerical simulations
\citep[e.g. using the ACE code,][]{Lohne2008}.

Our model is composed of a 1~M$_\odot$ star at the center and a debris
disc spanning 1 to 300 AU. The primordial mass surface density of
solids is assumed to be that of a MMSN: $\Sigma_0(r)=
(r/{\rm1\ AU})^{-1.5}$~\Mearth~AU$^{-2}$, with an initial size
distribution of solids proportional to $D^{-3.7}$, though the main
results presented below are independent of this choice. The minimum
size of solids in the cascade is set to 0.8 $\mu$m, which is the
blow-out size assuming a star of 1~L$_\odot$ and 1~M$_\odot$, and an
internal density of solids of 2700~kg~m$^{-3}$. Grains smaller than
this are immediately lost from the disc. We explore different maximum
diameters ($D_{\max}$) between 1-100 km. The disc is assumed to be
pre-stirred or stirred on a timescale much shorter than the age of the
system, i.e. initially having velocities high enough so collisions
between planetesimals are destructive and result in a collisional
cascade. This is accounted by setting the mean eccentricity ($e$) and
inclination ($I$) of the particles to be 0.05 and 1.4$^{\circ}$
($e/2$), respectively, which defines the relative velocities of the
particles. These velocities are calculated as $v_{\rm rel}=v_{\rm
  K}(1.25e^{2}+I^{2})^{1/2}$ \citep[valid for Rayleigh distributions
  of $e$ and $I$,][]{Lissauer1993, Wetherill1993}, where $v_{\rm K}$
is the Keplerian velocity on a circular orbit. Hence, the relative or
impact velocities are a 6\% of $v_{\rm K}$.

%% The catastrophic collision rate, $R_{\rm c}(D)$, is set by the density
%% of particles with enough incident energy to cause a catastrophic
%% collision, i.e. the number of particles larger than $X_{\rm c}D$, where 
%% \begin{equation}
%%   X_{\rm c}=(2Q_{\rm D}^{\star}/v_{\rm rel}^2)^{1/3}, \label{eq:Xc}
%% \end{equation}
%% and $Q_{\rm D}^{\star}$ is the disruption threshold of particles with
%% diameter $D$.

Furthermore, in our model destructive collision are only caused by
impactors with specific energies greater than the disruption threshold
or planetesimal strength ($Q_{\rm D}^{\star}$), which depends both on
the size and impact velocity. The disruption threshold has been
studied in laboratory experiments \citep[e.g.,][]{Fujiwara1989,
  Davis1990, Ryan1991} and with numerical simulations of colliding
basalt and icy bodies \citep[e.g.,][]{Benz1999}. It is well known that
for small bodies bound by cohesive binding forces, $Q_{\rm D}^{\star}$
decreases with size up to the size where self-gravity becomes
important, and then $Q_{\rm D}^{\star}$ increases with
size. Therefore, we assume the following prescription
\begin{equation}
  Q_{\rm D}^{\star}=\left[Q_{\rm D,s}\left(\frac{\rm D}{\rm 1\ m} \right)^{b_{\rm s}}+Q_{\rm D,g}\left(\frac{\rm D}{\rm 1\ m} \right)^{b_{\rm g}} \right]\left(\frac{v_{\rm rel}}{v_0}\right)^{1/2}, \label{eq:Qd}
\end{equation}
where $Q_{\rm D,s}$, $Q_{\rm D,g}$, $b_{\rm s}$ and $b_{\rm g}$ are
parameters that depend on the specific composition of solids in the
disc. The dependence on the relative or impact velocity is inspired by
the results from \cite{Stewart2009}. We use $Q_{\rm
  D,s}=500$~J~kg$^{-1}$, $Q_{\rm D,g}=0.03$~J~kg$^{-1}$, $b_{\rm
  s}=-0.37$, $b_{\rm g}=1.36$ and $v_0=3$~km~s$^{-1}$ values
consistent with Basalt in simulations from \cite{Benz1999}. The choice
of Basalt is not important for the results presented below. Using the
values estimated for planetesimals composed of ice from the same
study, we obtain similar results. Finally, we assume a
``redistribution function'' for the fragments created in a destructive
collision proportional to $D^{-3.5}$, with the largest fragment having
half the mass of the original disrupted body. The specific dependence
on $D$ does not change our results presented below.

\begin{figure}

  \begin{subfigure}
    
  \includegraphics[trim=0.0cm 0.5cm 1.0cm 0.0cm, clip=true,
    width=1.0\columnwidth]{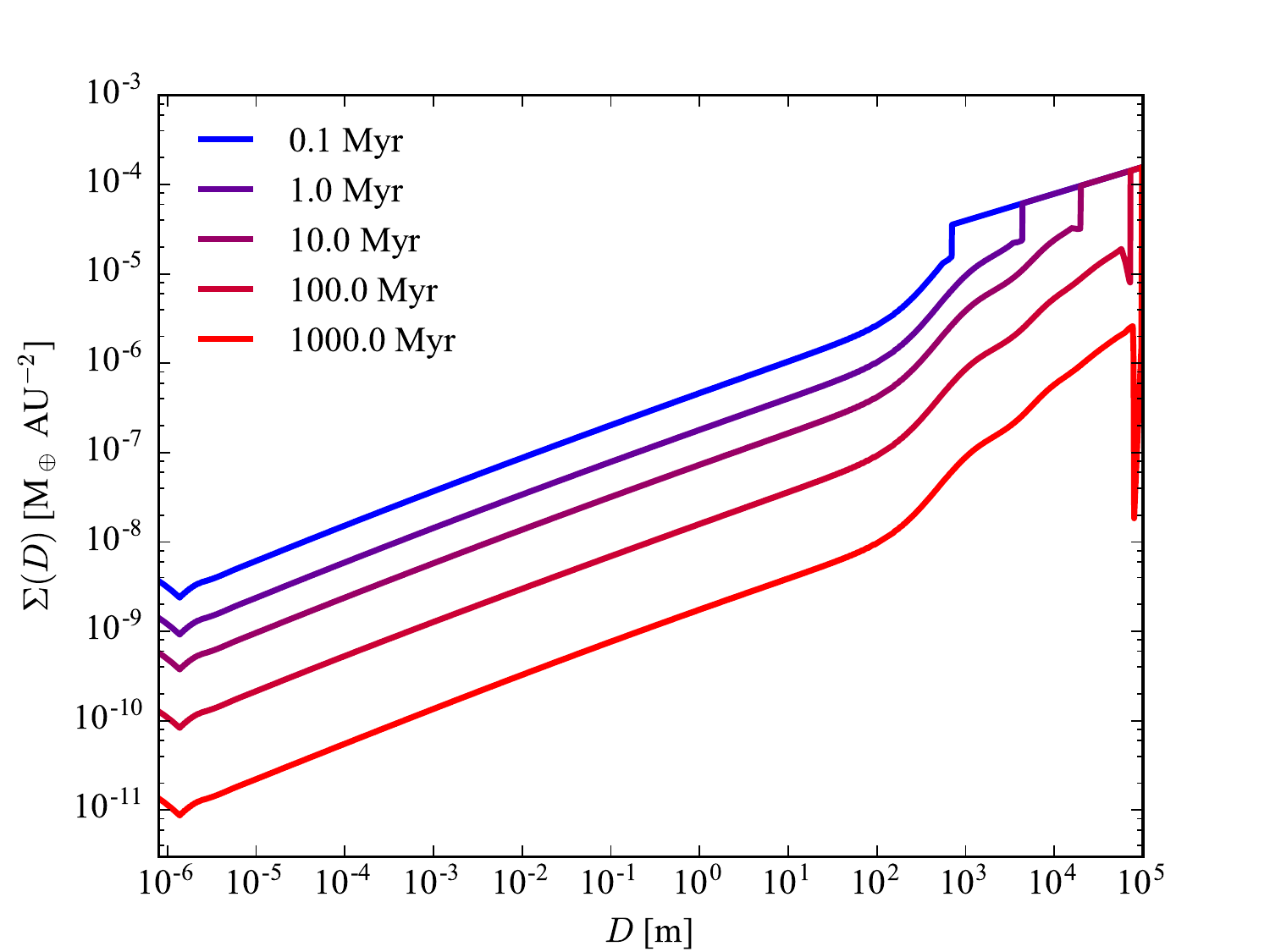}

  \end{subfigure}

  \begin{subfigure}
      
  \includegraphics[trim=0.0cm 0.0cm 1.0cm 1.0cm, clip=true,
    width=1.0\columnwidth]{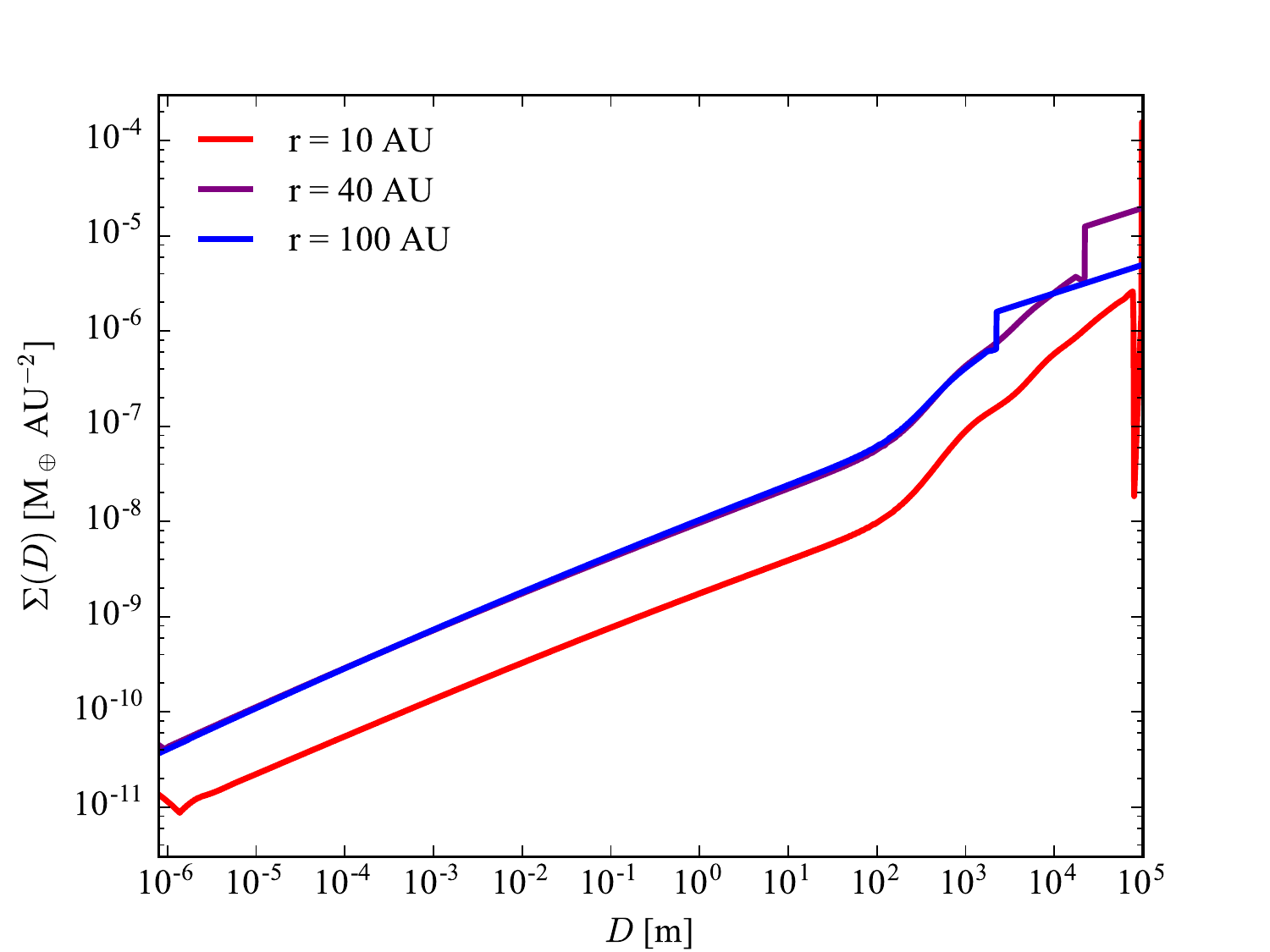}

  \end{subfigure}
  
  \caption{Mass surface density in each of the 3000 size bins spaced
    logarithmically, with $D_{\max}=100$~km and
    $\Sigma_0=$MMSN. Top: Size distribution at 10 AU for a system
    age ranging from 0.1 Myr to 1 Gyr. Bottom: Size distribution at
    10, 40 and 100 AU (red, purple and blue lines, respectively) for a
    system age of 1 Gyr.}
  \label{fig:Md}
\end{figure}

\begin{figure*}

  \begin{subfigure}%[]{0.5\columnwidth}

    \includegraphics[trim=0.0cm 1.5cm 17.0cm 0.5cm, clip=true,
      width=1.0\textwidth]{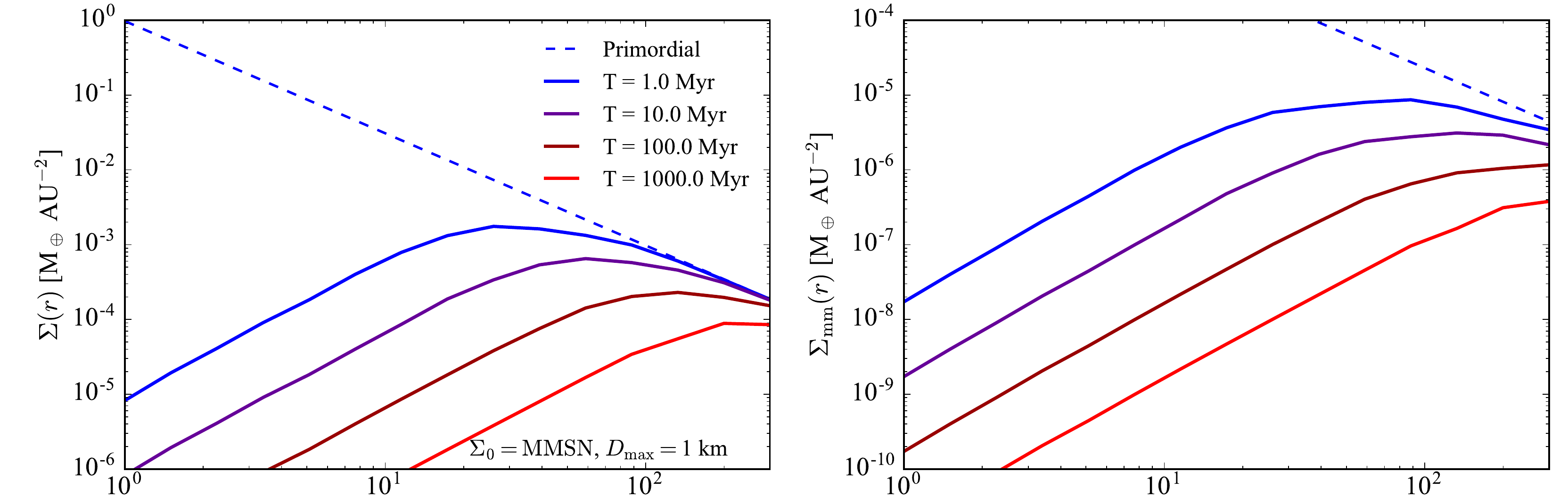}

  \end{subfigure}

  \begin{subfigure}%[]{0.5\columnwidth}

    \includegraphics[trim=0.0cm 1.5cm 17.0cm 0.5cm, clip=true,
      width=1.0\textwidth]{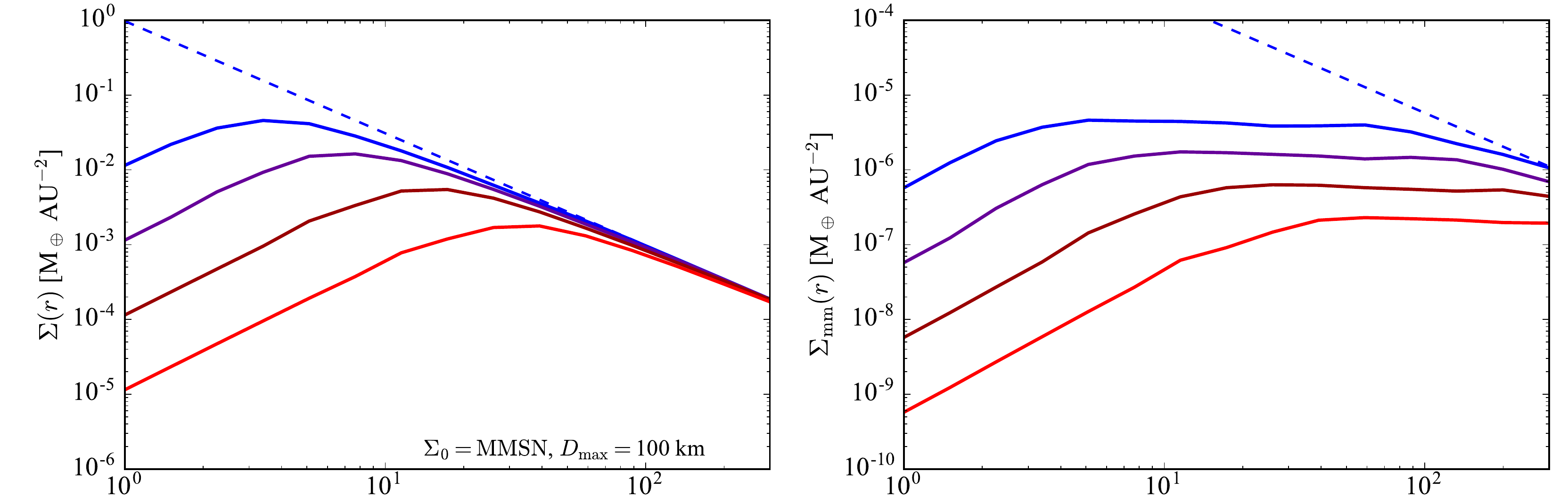}

  \end{subfigure}

    \begin{subfigure}%[]{0.5\columnwidth}

      \includegraphics[trim=0.0cm 0.0cm 17.0cm 0.5cm, clip=true,
        width=1.0\textwidth]{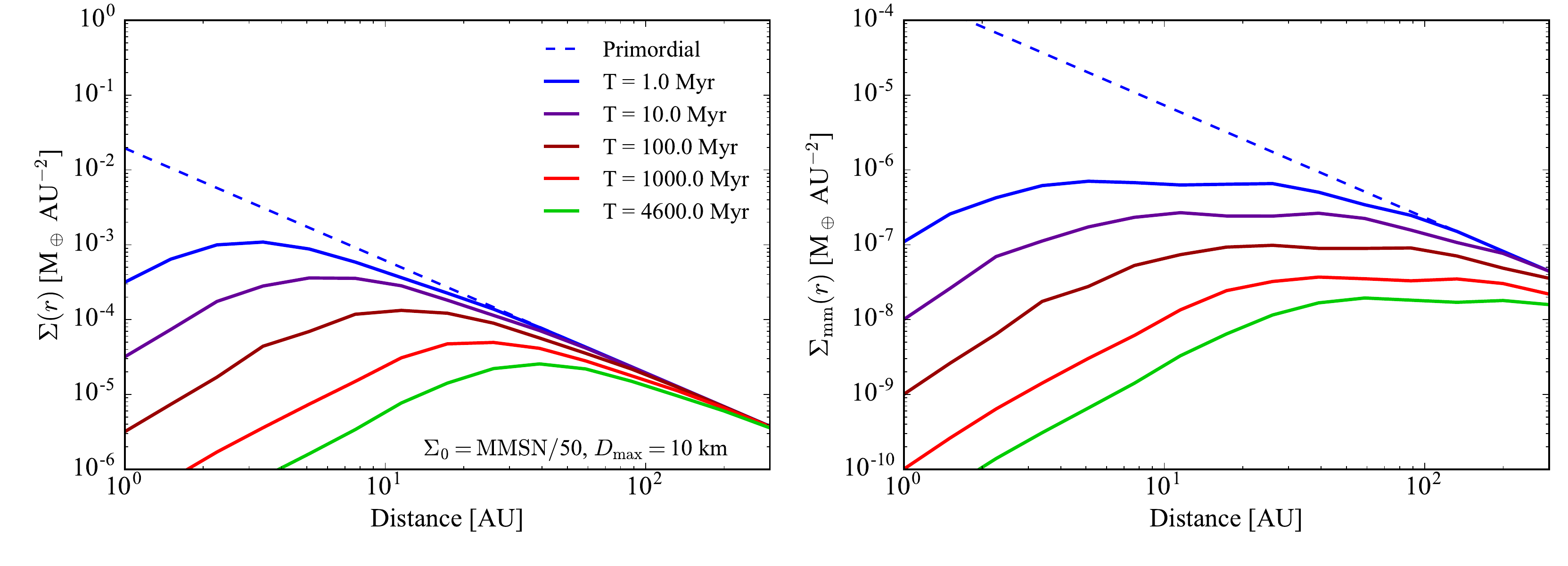}

  \end{subfigure}

  \caption{Total (left column) and millimetre-sized dust mass (right
    column) evolution of a disk from 1 to 300 AU. The different rows
    represent the evolution of a disc with the following parameters:
    (top) solar type star with a primordial surface density equal to a
    MMSN and a maximum planetesimal size of 1 km; (middle) solar type
    star with a primordial surface density equal to a MMSN and a
    maximum planetesimal size of 100 km; (bottom) central star of
    0.88~M$_\odot$ and 0.84~L$_\odot$ with a primordial surface
    density equal to 0.02 MMSN and a maximum planetesimal size of 5
    km. The colours represent 5 different ages: 1, 10, 100, 1000 and
    4600 Myr, varying from blue to red and green as time
    evolves. \label{fig:collevol}}
\end{figure*}

We divide the disc in different independent annuli, each one with a
total mass of $2\pi r \Delta r \Sigma_0(r)$, with $\Delta r=2er$,
which fixes the initial total mass in each radial bin. At a given
radius, we solve for the steady state size distribution by equating
the mass loss rate and gain in each size bin that is smaller than
$D_c$, the largest object that is in collisional equilibrium. The mass
in size bins larger than $D_c$ is held fixed to the primordial
distribution as they have lifetimes or collisional timescales longer
than the age of the system and have not had enough time to
significantly evolve. The timescale to reach quasi steady state or
damp perturbations is the same as the collisional timescale;
therefore, our quasi steady state assumption is valid for sizes
smaller than $D_c$.  To find the specific $D_c$ we solve for the
steady state size distribution varying $D_c$, until finding the
specific size bin with a lifetime equal to the age of the system (or
with a difference smaller than a 10\%). In the resulting size
distribution bins for planetesimals larger than $D_c$ retain their
original masses, while the masses in all smaller bins are anchored to
$D_c$ and their size distribution set by the collisional equilibrium
condition. As the system age increases, $D_c$ increases, and the size
distribution evolves, and thus, the total and mm-sized dust mass too.

%% maintaining the mass in the size distribution is adjusted to be
%% consistent with collisional equilibrium. For this, the size of the
%% largest object that is in collisional equilibrium $D_c$ must be
%% calculated in an iterative approach, solving for the steady state
%% distribution maintaining the mass fixed in size bins with $D>D_c$,
%% until we find $D_c$ such that its lifetime is equal to the age of the
%% system or with a difference smaller than a 10\%.

If $D_c>D_{\max}$, i.e. the lifetime of the biggest planetesimal is
shorter than the age of the system ($t_{\rm age}$), the mass in every bin is scaled as
\begin{equation}
  M(r, t, D)=M'_0(D) \frac{t_\mathrm{c}(0)}{t_{\rm age}}, \label{eq:timeevol}
\end{equation}
where $t_{\rm age}$ is the age of the system, and $M'_0$ is the mass
distribution in collisional equilibrium when $D_c=D_{\max}$, or when
the system had an age equal to the lifetime of the biggest
planetesimal, $t_\mathrm{c}(0)$. Equation \ref{eq:timeevol} is valid
if the mass loss rate is proportional to M$^2$, which is the case in
our models as the collisional lifetime is inversely proportional to
the mass in the cascade. The evolution of the surface density of
solids at 10 AU is illustrated in the top panel of Figure
\ref{fig:Md}. The main relevant feature of this evolution is that when
$D_c<D_{\max}$, the mass in the small size bins decreases more slowly
than it would when the entire size distribution is in equilibrium
($D_c=D_{\max}$).

%% To simulate the evolution of
%% the disc at different radii, we divide the disc in different
%% independent annuli, each one with a total mass of $2\pi r \Delta r
%% \Sigma(r)$, with $\Delta r=2er$, which fixes the initial total mass in
%% each radial bin.

%% Then, in an iterative approach we solve for the steady state
%% distribution maintaining the mass fixed in size bins with $D$ bigger
%% than an initial guess of $D^{\rm t}$ (the maximum size in collisional
%% equilibrium), which is then changed iterating until the collisional
%% lifetime of $D^{\rm t}$ is equal to the age of the system or with a
%% difference smaller than a 10\%. If the lifetime of the biggest
%% planetesimal is shorter than the age of the system, then only that
%% size bin is held fixed to compute the size distribution in
%% equilibrium. Finally, the mass in every bin is scaled as
%% \begin{equation}
%%   M(r, t, D)=\frac{M'_0(D)}{1+t/t_\mathrm{c}(0)}, \label{eq:timeevol},
%% \end{equation}
%% where $M'_0$ is the mass distribution in collisional equilibrium at
%% $t=0$ and $t_\mathrm{c}(0)$ is the collisional lifetime of the largest
%% planetesimal in the disc at $t=0$. Equation \ref{eq:timeevol} is valid
%% if the mass loss rate is proportional to M$^2$, which is the case in
%% our models as the collisional lifetime is inversely proportional to
%% the mass in the cascade.

In Figure \ref{fig:collevol} we present the evolution of three discs
varying $D_{\max}$ from 1 to 100 km (top and middle), and changing the
stellar mass and luminosity together with $\Sigma_0$ and $D_{\max}$ to
fit 61~Vir disc properties (bottom panel), i.e. its surface density of
mass in mm-sized grains and disc inner edge (see
Sec. \ref{sec:collmodel61vir}). The surface density of the total mass
in solids ($\Sigma$, left column) evolves with time similarly to
analytic models \citep[e.g.,][]{Wyatt2007collisionalcascade},
increasing with distance as expected up to the characteristic radius,
$r_c$, at which the largest planetesimal in the disc has a lifetime
equal to the age of the system, i.e. $t_\mathrm{c}(0)=t_{\rm age}$,
and from there decreasing with radius as $\Sigma_0(r)$. This radius
depends on the initial total solid mass and on $D_{\max}$ as the three
panels in the first column show, with $r_c$ being smaller for larger
$D_{\max}$ or lower initial mass as the rate of collisions is
reduced. The surface density of mass in mm-sized grains
($\Sigma_\mathrm{mm}$, right column) behaves in a way similar to the
optical depth described in \cite{Schuppler2016} and fractional
luminosity in \cite{Geiler2017}, mimicking $\Sigma(r)$ for $r<r_c$,
but considerably flatter compared to $\Sigma(r)$ at $r>r_c$. This is
because $D_c<D_{\max}$ and $D_c$ decreases with $r$ outside $r_c$, so
$\Sigma_{\rm mm}$ is less depleted for larger $r$. The net effect is
that $\Sigma_{\rm mm}(r)$ is almost constant, even though $\Sigma(r)$
decreases with $r$.

%% The surface density of millimetre grains ($\Sigma_\mathrm{mm}$, right
%% column) behaves in a way similar to the optical depth described in
%% \cite{Schuppler2016}, mimicking the total surface density for $r<r_c$,
%% but considerably depleted compared to the total mass at $r>r_c$.

To illustrate the differences in the evolved size distribution at
different radii, in the bottom panel of Figure \ref{fig:Md} we compare
the size distribution at 1 Gyr with $D_{\max}=100$~km and $r=10$, 40
and 100 AU. At 10 AU (red line) the disc evolves fast as relative
velocities are higher and all the size bins are in collisional
equilibrium. At 40 AU (purple line) relative velocities are slower,
hence $D_c\sim20$~km and only smaller bodies are in collisional
equilibrium; therefore, the mass in small bodies is highly depleted
compared to the primordial, while the total mass in solids has not
decreased significantly. At 100 AU (blue line) relative velocities are
even slower, $D_c\sim2$~km and the mass in small bodies is less
depleted compared to the primordial than at 40 AU. Even though the
surface density of the total mass in solids at $t=0$ and 1~Gyr is
higher at 40 AU than at 100 AU, the mass surface density in solids
smaller than 1 km is approximately the same at both radii after 1
Gyr. This causes the slope of $\Sigma_\mathrm{mm}$ to flatten out and
be almost constant at large radii where $t_\mathrm{c}(0)>t_{\rm age}$
as mentioned above. We also observe a very similar evolution for the
vertical optical depth in the disc, consistent with
\cite{Schuppler2016}.

This behaviour that makes $\Sigma_{\rm mm}$ to be almost flat can be
understood analytically if we consider a planetesimal strength
approximated by two broken power laws and a continuous size
distribution with three regimes: i) small bodies in collisional
equilibrium with a size distribution proportional to $D^{-q_1}$; ii)
large bodies with gravity dominated strengths in collisional
equilibrium with a size distribution proportional to $D^{-q_2}$; and
iii) largest planetesimals with lifetimes longer than the age of the
system that conserve their primordial size distribution proportional
to $D^{-q_3}$. The value of $q_1$ and $q_2$ are strictly related to
the dependence on $D$ for $Q_{\rm D}^{\star}$, with \citep{Durda1997,
  OBrien2003}
\begin{equation}
q_{i}=\frac{21+b_i}{6+b_i},
\end{equation}
where $b_i$ is the slope or exponent of $Q_d^{\star}$ in the strength
or gravity-dominated regime. Therefore, assuming reasonable values for
$b_s$, $b_g$ and $q_3$ we can find an analytic expression for the size
distribution at different radii \citep[e.g.,][]{Lohne2008}. Moreover,
assuming an initial surface density or mass distribution in the disc,
we can derive an expression for the fractional luminosity as a
function of radius, as shown by \cite{Shannon2011} (Equation A10
therein) and \cite{Geiler2017} (Equation A11 therein). As the
fractional luminosity is proportional to the Surface density of small
grains, we can rewrite Equation A10 in \cite{Shannon2011} to find
\begin{equation}
  \Sigma_{\rm mm} (r) \propto \left[r^2\Sigma_0(r)\right]^{\frac{2+k_2-k_2q_2}{2+q_2-q_3+k_2-k_2q_2}} r^{-2+\frac{(19+2q_2)(q_2-q_3)}{6(2+q_2-q_3+k_2-k_2q_2)}}, \label{eq:shannon0}
\end{equation}
where $\Sigma_0(r)$ is the primordial total surface density of solids,
and $k_2$ is equal to $\frac{6-q_2}{q_2-1}$ and represents the size
scaling of the minimum impactor size to cause a catastrophic
collision. The expression above is only valid when $D_c$ is less than
$D_{\max}$, but large enough so it is in the gravity dominated regime
($D_c\gtrsim100$~m). Assuming $q_1$=3.6, $q_2$=3.0, $k_2=1.5$ (values
consistent with $b_s$ and $b_g$ used above), $q_3$=3.7 and
$\Sigma_0(r)=\Sigma_0 (r/1\ {\rm AU})^{\alpha}$ we find
\begin{equation}
  \Sigma_{\rm mm} (r) \propto r^{0.6\alpha+0.9}. \label{eq:shannon}
\end{equation}
Therefore, for $\alpha=-1.5$, $\Sigma_{\rm mm}$ is independent of
radius, which matches with our more detailed numerical
simulation. Moreover, the flatter $\Sigma_{\rm mm}$ in the evolved
size distribution compared to the primordial distribution is
independent of $\alpha$ as Equation \ref{eq:shannon} shows; although a
steeper primordial surface density of solids decreasing with radius
would result in a steeper surface density of millimetre sized grains
with a slope of $0.6\alpha+0.9$. For $q_3=3.5$ and 3.9 we still find a
flat slope for $\Sigma_{\rm mm}$ of -0.3 and 0.2, respectively. From
the results in our simulations we can estimate the dependence of
$\Sigma_{\rm mm}$ on $t,D_{\max}$ and $\Sigma_0$, by assuming a power
law dependence and fitting it to our numerical results. Coupling these
with the dependence on $r$ from Equation \ref{eq:shannon} (only valid
for $q_1$=3.6, $q_2$=3.0 and $q_3$=3.7), we find
\begin{equation}
  \begin{split}
    \Sigma_{\rm mm}(r>r_c) \approx 2 \left(\frac{r}{1\ {\rm AU}}\right)^{0.6\alpha+0.9}
    \left(\frac{t}{1\ \mathrm{Gyr}}\right)^{-0.4}
    \left(\frac{D_{\max}}{100\ \mathrm{km}}\right)^{-0.1}\\
    \left(\frac{\Sigma_0}{1\ \mathrm{MMSN}}\right)^{0.6} 
    {\rm M}_{\oplus}~{\rm AU}^{-2}, \label{eq:Sigmamm}
  \end{split}
\end{equation}
where $\Sigma_0$ is the initial surface density of solids at 1 AU in
units of the MMSN. The factor 2 and the exponents of $-0.4$, $-0.1$
and $0.6$ are the results from a fit to the numerical simulations. Equation
\ref{eq:Sigmamm} is only valid for $r>r_c$ and
$D_c\gtrsim100$~m. Using Equation A5 in \cite{Shannon2011} we can also
estimate how $r_c$ varies with time and $\Sigma_0$. Moreover, from our
simulations we can derive a dependence on $D_{\max}$ fitting a power law. We find
\begin{equation}
  r_c \approx 4 \left(\frac{t}{1\ \mathrm{Myr}}\right)^{\frac{1}{-\alpha+1.5}}
  \left(\frac{\Sigma_0}{1\ \mathrm{MMSN}}\right)^{\frac{1}{-\alpha+1.5}}
  \left(\frac{D_{\max}}{100\ \mathrm{km}}\right)^{-0.5} \ \mathrm{AU}, \label{eq:rc}
\end{equation}

Assuming a specific dependence of planetesimal strength on size,
equations A10 from \cite{Shannon2011}, and \ref{eq:Sigmamm} and
\ref{eq:rc} from this work, together with the disc model presented
above, can be used to retrieve the primordial radial distribution of
solids from ALMA observations of extended discs if the biggest
planetesimals in the disc still conserve their primordial size
distribution. Moreover, they can be used to constrain the initial
total mass in the disc and maximum planetesimal size. So far there is
no evidence of extended debris discs at millimetre wavelengths with a
steep slope decreasing with radius \citep[or non consistent with being
  flat, e.g.,][]{Booth2016}; however, even with ALMA (the most
sensitive instrument at millimetre wavelengths) the detection and
study of extended debris discs is only possible around a few of the
brightest systems.

It is worth noting that the maximum planetesimal size in a disc could
vary with radius by orders of magnitude as the growth timescales for
planetesimals are a steep function of radius and the surface density
in solids \citep[e.g.,][]{Kenyon2008}. Moreover, stirring could have
stopped the growth at different epochs for different radii. Although
in our models we assume that the maximum planetesimal size is
independent of radius, our prediction for $\Sigma_{\rm mm}(r>r_c)$ in
Equation \ref{eq:Sigmamm} is not very sensitive to
$D_{\max}$. Therefore, our predictions are reasonably valid even if
the maximum planetesimal size decreases with radius (as expected in
planet formation models). This is already illustrated in Figure
\ref{fig:collevol}. If we consider a disc with $D_{\max}$ decreasing
from 100 to 1 km between 40-300 AU, then the resulting $\Sigma_{\rm
  mm}(r)$ at 1 Gyr would be almost the same as the red line in the
middle right panel on that Figure, because $\Sigma_{\rm mm}(300\ {\rm
  AU})$ increases only by a factor of 2 when decreasing $D_{\max}$
from 100 to 1 km. This is due to two opposite effects: 1) for a
constant total mass in solids, reducing $D_{\max}$ increases the mass
in millimetre sized dust; and 2) reducing $D_{\max}$ makes the
collisional evolution faster which reduces the mass in every bin in
collisional equilibrium. A similar effect would be present at $r<r_c$
making the surface density slope flatter. The maximum planetesimal
size is only significantly important to determine $r_c$. The opposite
scenario, and less likely, in which $D_{\max}$ increases with radius
would result in a slightly steeper slope for both $r<r_c$ and $r>r_c$.

%% To illustrate these two possible cases we compute the collisional
%% evolution for a disc in which $D_{\max}$ varies with radius like
%% power law from 100 km at 1 AU and 1 km at 300 AU, and the vice
%% versa. These are ilustrated in Figure \ref{}

Other differences relative to our assumptions could also change the
slope of the millimetre surface density, such as the epoch of stirring
(in our simulations we consider a pre-stirred disc), or the mean
eccentricity and inclination of particles in the disc, or even the
disruption threshold of planetesimals and dust if their composition
varies with radius. For example, a different $Q_{\rm D}^{\star}$ would
modify the size distribution, changing the slope of the predicted
millimetre surface density as Equation \ref{eq:shannon0} shows.

%% For example, HR~8799 was
%% constrained to have a millimetre surface density slope of $-1\pm0.4$
%% which could, for example, fit the slope of the millimetre surface
%% density derived in other discs such as
%% \citep[e.g.,][]{Booth2016}.

%% Otherwise, the mass in the bin is
%% set by the collisional equilibrium. calculated If the collisional
%% timescale of the largest planetesimal in the disc is shorter than the
%% age of the system, then corresponding mantain fixed the mass on each
%% bin with

\subsection{Application to 61~Vir}
\label{sec:collmodel61vir}

In Sec. \ref{sec:model} we find that the observations in the
millimetre are best fitted with a disc extending to $\sim150$ AU, an
integrated flux of $3.7\pm1.2$~mJy and a flat surface density
distribution, equivalent to a dust mass of
$\sim2\times10^{-8}$~M$_\oplus$~AU$^{-2}$. In addition, the minimum
radius derived from a best fit model of a collisionally evolved disc
to the Herschel observations is $\sim40$~AU. Using the same model for
the collisional evolution of a disc described above \citep[replacing
  the stellar mass and luminosity with 0.88 M$_\odot$ and 0.84
  L$_\odot$,][]{Sousa2008, Wyatt2012} we find a best match with a
primordial surface density between 20-100 times less dense than the
MMSN and a maximum planetesimal size between $5-20$ km. These two
parameters determine that $t_c(0)=4.6$~Gyr at $\sim40$~AU and
$\Sigma_\mathrm{mm}(r)\sim2\times10^{-8}$~M$_\oplus$~AU$^{-2}$ for
$r>40$~AU.

The need for a low primordial surface density and a maximum
planetesimal size of 10~km is due to the low mass in millimetre
grains, which scales roughly as $D_{\max}^{-0.1}\Sigma_0^{0.6}$ (see
Equation \ref{eq:Sigmamm}), together with a large $r_c$, that scales
roughly as $D_{\max}^{-0.5}\Sigma_0^{0.33}$ (see Equation
\ref{eq:rc}). Therefore, we need a very low $\Sigma_0$ to fit the
millimetre surface density and a low $D_{\max}$ to have $r_c\sim40$~AU
given the low $\Sigma_0$. From the size distribution we can also
determine a vertical optical depth of $2\times10^{-4}$, a few times
higher than the optical depth from Herschel observations and SED
fitting, but still consistent considering all the assumptions made in
both modelling efforts. For example, a more detailed treatment of
radiation pressure could change the value of the optical depth by a
factor of a few. The derived maximum planetesimal size and primordial
surface density go in the same direction as the ones from
\cite{Wyatt2012}: the primordial surface density of solids in the disc
was much lower compared to the MMSN and with a maximum planetesimal
size not much larger than 10 km.

%% and surface density
%% $\sim2\times10^{-8}$~M$_\oplus$~AU$^{-2}$.

%% Using the same model we find that in order to reproduce a flat surface densit

\section{Discussion}
\label{discussion}

\subsection{A depleted broad disc of planetesimals}

In Sec. \ref{sec:model} we found that the debris disc in 61~Vir is
broad, extending from 30 to 150 AU or larger radii. If the emission
were concentrated in a few $\lesssim$20 AU wide rings of planetesimals
the disc would have been detected above 3 sigma in the ALMA
map. Moreover, the 2.2$\sigma$ difference between the flux measured by
SCUBA2 and ALMA is indicative that there is flux loss in the
reconstructed ALMA image due to the disc emission being mostly in
structures on scales larger than $6\arcsec$ (50 AU); and thus not
recoverable by the range of baselines in the ALMA data. This was
corroborated using simulated observations of different broad disc
models. Therefore, we conclude that the planetesimal disc must be
extended with a wide range of semi-major axes. A different scenario
with a population of highly eccentric planetesimals with a small range
of semi-major axes is discarded as the derived surface density is
flatter than expected in a scattered disc scenario
\citep[e.g.,][]{Duncan1997} and while collisional erosion can flatten
this distribution by preferentially eroding the inner regions this
cannot completely erase the density enhancement at the inner edge of
the disk \citep{Wyatt2010}.

%% We can discard o $r^{-3.5}$ \citep{Duncan1997} either with a population of highly eccentric planetesimals
%% or with a wide range of semi-major axes.

The inner edge of the disc could be defined by the collisional
evolution that has been ongoing for 4.6 Gyr as assumed in
Sec. \ref{sec:collmodel}, or instead the disc could have been
truncated by a yet unseen planet. In the first scenario, the observed
inner edge of the disc (30-40 AU) can be explained by a maximum
planetesimal size of only about 10 km and primordial surface density
of solids 50 times lower than a MMSN. One explanation for why the
planetesimals did not grow to larger sizes could be the low surface
density of solids which slows down the growth timescales
\citep{Kenyon2008}, but could also be because the planetesimals were
stirred by a planet closer in hindering their growth.

%% Although
%% numerical models show that it is relatively easy for planetesimals
%% between 30-150 AU to grow up to $\sim$1000 km in size before the disc
%% is stirred by the same protoplanets igniting a collisional cascade
%% \citep[e.g.,][]{Kenyon2008}. However, the same studies show that the
%% growth timescales depend on the initial disc mass, which if too low
%% could have hindered the growth of bigger planetesimals. For example,
%% the growth timescales could have not been fast enough for
%% planetesimals to grow up to 1000 km in size before being stirred by a
%% planet closer in.

In the second scenario, in which the inner edge of the disc has been
truncated by a planet, the maximum planetesimal size is no longer
constrained to be of the order of $\sim10$~km. However, even if we
consider a maximum planetesimal size of 1000 km, the mass of the
primordial disc still needs to be a factor of $\sim10$ lower compared
to a MMSN in order to fit the flat surface density of millimetre
grains derived in this paper and the Herschel observations
\citep{Wyatt2012}. This depletion could arise from the protoplanetary
disc phase if the disc had a low mass, or a low efficiency of
planetesimal formation, or due to radial drift of solid particles
during that gas rich phase which concentrated most of the solid mass
in the inner regions \citep{Whipple1973, Weidenschilling1977drag}. The
radial drift of solids could have also contributed to an in situ
formation of the 2-3 planets found within 1 AU of the Star
\citep[e.g.,][]{Hansen2012}.

A variant on the second scenario involves the 30 AU truncation radius
being caused by a planet which is no longer present. For example, if
the close-in planets formed further out (just inside 30au) and then
migrated to their current location accreting and scattering
planetesimals on their way in, the early evolution of these close-in
planets could be responsible for both the truncation of the outer disc
and its stirring \citep[e.g.,][]{Alibert2006, Terquem2007,
  Kennedy2008hotsuperearth, Payne2009, IdaLin2010}.

%% i.e. lying at the right edge of the blue-white region
%% in Figure \ref{fig:mplt},

%% As the three planets in the sy   if planetesimals were not initially in eccentric
%% orbits that could have ignited a collisional cascade, e.g. via planet
%% disc interactions if planets migrated in through the disc, these must
%% have been stirred by other mechanism up to 150 AU, different to secular
%% interaction with the three known planets
\subsection{Stirring by a yet unseen planet}

\begin{figure}
  \includegraphics[trim=0.5cm 0.0cm 2.0cm 0.0cm, clip=true,
    width=1.0\columnwidth]{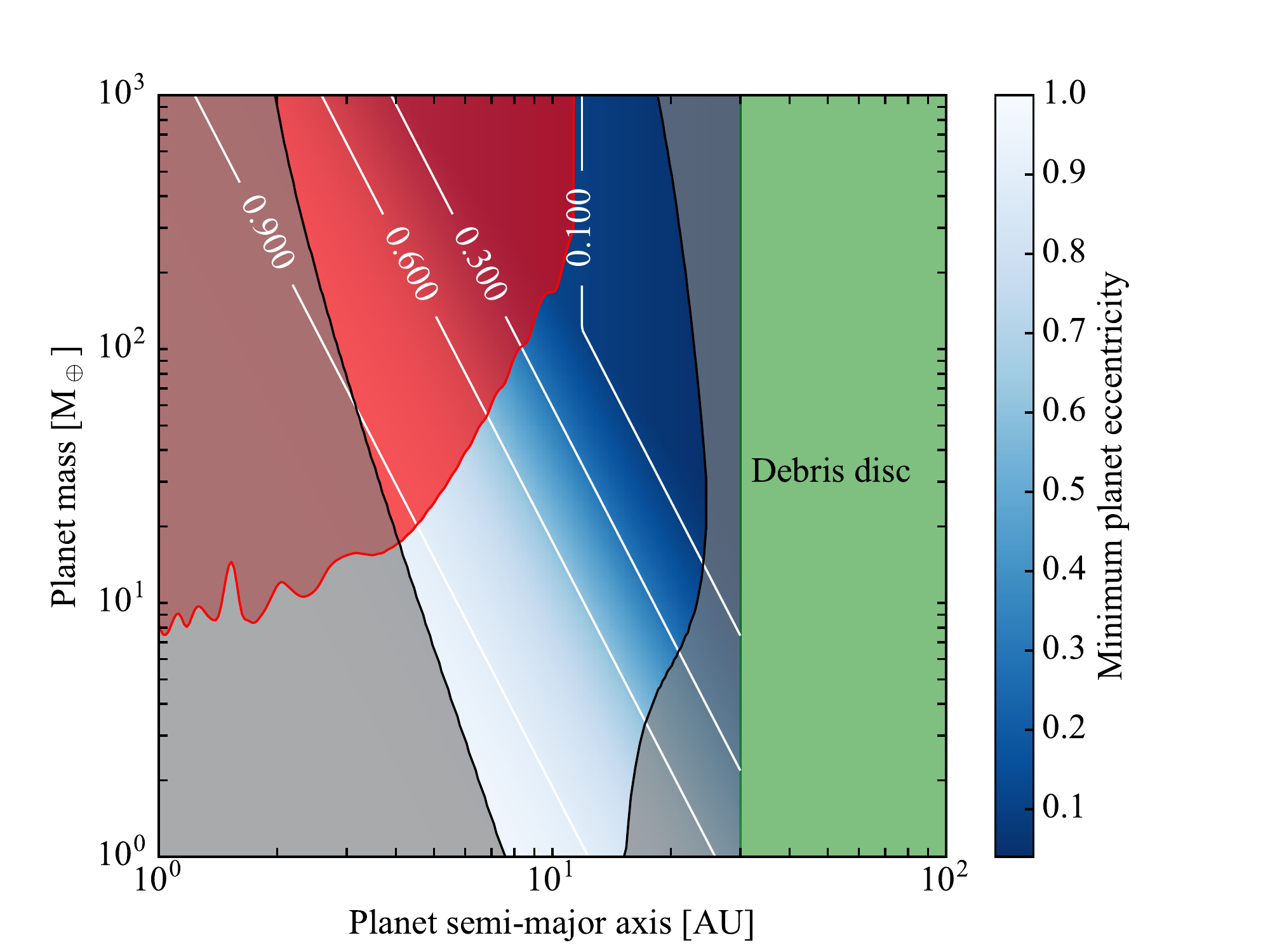}
  \caption{Allowed masses and semi-major axes for a putative planet
    that stirred the 61 Vir debris disc out to 150 AU, in a timescale
    shorter than 4.6 Gyr, and forcing an eccentricity higher than
    0.01. The blue colour map and white contours represent the minimum
    eccentricity for a given planet mass and semi-major axis. The green
    shaded region on the right is excluded as the planet would overlay
    with the inner edge of the disc at 30~AU. The grey shaded region
    is excluded as those planets would get closer than 5 Hill radii to
    61 Vir c or to the inner edge of the disc. Finally the red region
    in the top left corner is excluded from upper limits based on RV
    data.}
  \label{fig:mplt}
  
\end{figure}

If 61 Vir b and c formed in situ, then something else must have
stirred the disc as these are too far in and not massive enough to
have stirred the disc at large radii within 4.6 Gyr
\citep{Wyatt2012}. Hence, we propose that an unseen planet at a larger
distance and within the 30 AU disc inner edge stirred the
disc. Similar to \cite{Moor2015herschel}, using Eq. 6 from
\cite{Mustill2009} (valid for planets with eccentricities $\lesssim
0.3$) we can derive lower limits on the eccentricity of such a planet
depending on its semi-major axis and mass so the timescale of stirring
is shorter than the age of the system. Moreover, the eccentricity
imposed on the planetesimals ($e_f$) must be higher than a certain
value so that their relatives velocities are high enough to cause
destructive collisions
($v_\mathrm{rel,max}\sim2e_\mathrm{f}v_\mathrm{K}$). Here we impose
that the forced eccentricity \citep[Equation 8 in][]{Mustill2009} must
be higher than 0.01 so planetesimals of 5 km diameter undergo
destructive collisions with planetesimals of the same size at 150
AU. This is illustrated in Figure \ref{fig:mplt}. The minimum
eccentricity decreases with increasing semi-major axis and planet mass
as the timescale for stirring is held fixed at 4.6 Gyr. The forced
eccentricity must be $>0.01$, which results in a kink in the 0.1
contour (because $e_f$ is independent of mass). All other contours are
set by the stirring time set equal to the age of the system.

%% i.e. $X_{\rm c}=1$ using Equations \ref{eq:Xc} and \ref{eq:Qd}

%% unless it results in
%% a forced eccentricity lower than 0.01, in which case we fixed it to
%% 0.01 which is independent of the planet mass and results in a stirring
%% timescale shorter than the age of the system (e.g. white vertical
%% contour).

We can add additional constrains if we require planets with a
pericenter that does not get closer than 5 mutual Hill radii
\citep[see Eq. 9 in][]{Pearce2014} from the apocentre of 61~Vir~c
(a=0.22 AU, e=0.14), i.e
\begin{equation}
  a_{\rm plt}(1-e)-5 R_{\rm H,q} > a_{\rm c} (1+e_{\rm c}), 
\end{equation}
where $R_{\rm H,q}$ is the Hill radii at pericentre and $a_{\rm c}$
and $e_{\rm c}$ are the semi-major axis and eccentricity of
61~Vir~c. In addition, the apocentre of the hypothetical planet d has
to be such that it does not get closer than 5 Hill radii to the disc
inner edge at $\sim 30$~AU. These two constraints exclude the grey
shaded area. Because lower mass planets have higher eccentricity, the
maximum semi-major axis decreases with decreasing planet mass for
M$_{\rm plt}\lesssim10$~M$_\oplus$, but also decreases with increasing
planet mass as the $R_{\rm H,q}$ gets larger. Finally, using upper
limits from RV data from HARPS we can exclude planets more massive
than the red line \citep{Wyatt2012, Kennedy2015superearths}.

With these limits on M$_{\rm plt}$ and $a_{\rm plt}$ we can conclude
that if an unseen planet interior to the debris disc is responsible
for stirring the planetesimals up to 150 AU, and has an eccentricity
lower than 0.1, then it must be more massive than 10 \Mearth\ and have
a semi-major axis between 10-20 AU. Less massive planets and closer in
($a_{\rm plt}=4-20$~AU) could have stirred the disc, but with
$e\gg0.1$. For the allowed combinations of $M_\mathrm{plt}$ and
$a_\mathrm{plt}$ even a highly eccentric planet will not induce an
eccentricity higher than the observed on 61 Vir b and c or cause close
encounters \citep[see Figure 5 in][]{Read2016}. Moreover, an eccentric
planet will impose an eccentricity on the disk which may be detectable
by imaging of the disc \citep{Wyatt1999}. While there is no evidence
of any asymmetry, the constraints are weak, both because the imaging
is in the far-IR where the transition from pericentre to apocentre
glow occurs \citep{Pan2016}, and because the disc would look symmetric
if the pericentre is aligned with the minor axis of the disc projected
in the sky.

The equations used to derive the minimum eccentricity are only valid
for $e_{\rm plt}\lesssim 0.3$. Planet eccentricities higher than 0.3
could be overestimated as the predicted stirring timescales are longer
than expected for $e>0.3$ \citep[see Figure 1
  in][]{Mustill2009}. Therefore, the lower limits presented in Figure
\ref{fig:mplt} are only representative of the constraints expected for
$e_{\rm plt}\lesssim 0.3$.

%% ...

%% - radial drift
%% - stirring stopped the growth
%% - the low primordial disc could have determained the low maximum size (kenyon and Bromley 2012?)

%% is further than 150 AU, with a continuous
%% distribution from 30

%% No massive planets at large radii?

\subsection{Background sources}
\label{dis:bgal}

As noted before by \cite{Wyatt2012}, none of the detected compact
sources (B1 and B2) are co-moving with 61~Vir, therefore, we can
assume these are background objects. B1 together with B3 and its
northern counterpart (which lies outside the ALMA primary beam) are
probably related to an active galactic nucleus
\citep[AGN,][]{Condon1998}. B1 at the center is consistent with
compact emission from dust heated by an AGN, or with flat-spectrum
synchrotron emission typical of a radio galaxy core, or with dust
associated with a nuclear starburst, or some combination of these
possibilities.  On the other hand, the two lobes are consistent with
synchrotron emission that we do not expect to detect in the
sub-millimetre (sub-mm), given their steep spectra.

%% unresolved galaxy detected in thermal emission
%% with ALMA.  with two lobes detected at 1.4 and 5 GHz with the VLA. The
%% two lobes can be associated to synchrotron emission from two jets
%% arising from an active galactic nucleus \citep[AGN,][]{Condon1998} at
%% the center of the galaxy (B1).

The most northern source detected by ALMA (B2) was previously detected
by Herschel from 160 to 500 $\mu$m and is resolved by the ALMA
synthetic beam with a size of $\sim2\arcsec$. This is larger than
expected for the z$\sim$1-3 sub-mm galaxy population
\citep[e.g.,]{Smail1997}, where ALMA has measured typical sub-mm sizes
of $\ll1\arcsec$ \citep[e.g.,][]{Ikarashi2015, Simpson2015}, slightly
smaller than their typical radio sizes \citep{Biggs2011}, but is
consistent with a dusty starburst at a rather lower redshift, as we
would suspect from its relatively blue PACS/SPIRE SED.

%% This is consistent with
%% cold thermal emission from a background galaxy. On the other hand, B1
%% together with B3 and its northern counterpart (which lies outside the
%% ALMA primary beam) are probably related to an unresolved galaxy
%% detected in thermal emission with ALMA, with two lobes detected at 1.4
%% and 5 GHz with the VLA. The two lobes can be associated to synchrotron
%% emission from two jets arising from an active galactic nucleus
%% \citep[AGN,][]{Condon1998} at the center of the galaxy (B1).

%% previously detected by the VLA at 1.4
%% GHz as an extended source with a FWHM of $33\arcsec$ along the major
%% axis with a PA of $\sim4^{\circ}$ and a FWHM of $<18\arcsec$ along its
%% minor axis \citep{Condon1998}. The other three background sources
%% reported in \cite{Wyatt2012} and detected in the Herschel and VLA data
%% lie outside the field of view of these ALMA observations. In Figure
%% \ref{fig:} we summarize the position of 5 background objects with
%% respect to the 61 Vir at 2015 April.

\section{Summary and Conclusions}
\label{conclusions}

We have presented the first resolved millimetre study of 61 Vir, a
planetary system with two confirmed RV planets within 1 AU and a
debris disc at tens of AU. Combining ALMA and SCUBA2/JCMT observations
we found that at 0.86~mm the total disc emission is $3.7\pm1.2$~mJy,
the disc extends from 30 to at least 150 AU, and has surface density
exponent of millimetre grains of $0.1^{+1.1}_{-0.8}$. This implies
that the parent planetesimal disc is broad with a wide range of
semi-major axes. The alternative scenario of a highly scattered disc
with planetesimals with a common pericentre is discarded given the
constraints on the surface brightness of the disc. No CO gas emission
was detected in the disc, although even if planetesimals are rich in
CO and releasing gas through collisions, we predict that any emission
should be below our detection limit.

We developed a full disc collisional evolution model based on previous
numerical work that can reproduce some of the results obtained in more
detailed simulations, but in a much more computationally efficient
approach. These models predict that the surface density of millimetre
grains and optical depth radial profiles do not necessary match with
the surface density of the parent bodies, tending to be flatter in
regions of the disc where the age of the system is shorter than the
collisional lifetime of the biggest planetesimals. This can be used to
constrain the primordial surface density distribution of solids and
maximum planetesimal size for extended discs for reasonable
assumptions on the eccentricity, inclination and strength of
planetesimals. For example, with this model we can reproduce the
observations if 61~Vir debris disc started with a surface density
$\sim50$ times more depleted in solids compared to a MMSN, and with
planetesimals that did not grow more than 5-20 km in size so the disc
is collisionally depleted at $r<40$~AU. However, these conclusions are
based on the assumption that the inner edge of the observed disc is
set by the collisional evolution of the disc. If instead the inner
edge is set by other mechanism, e.g. planet-disc interaction, then the
maximum planetesimal size is no longer constrained, but the primordial
surface density would still need to be depleted by a factor of
$\sim10$ compared to the MMSN.

Finally we discussed and constrained the mass, semi-major axis and
eccentricity of a planet stirring the disc located between the known
RV planets and the inner edge of the disc. We found that in order to
have stirred the disc out to 150 AU, the planet must be more massive
than 10~$M_\oplus$ and a semi-major axis between 10 and 20 AU if it
has an eccentricity lower than 0.1. Otherwise, for higher
eccentricities it could have a lower mass and a semi-major axis
between 4 and 20 AU.

\section*{Acknowledgements}

We thank Pablo Roman and Simon Casassus for providing us the tool
uvsim to simulate model visibilities. We also thank Matthew J. Read
for useful discussion. This paper makes use of the following ALMA
data: ADS/JAO.ALMA\#2013.1.00359.S. ALMA is a partnership of ESO
(representing its member states), NSF (USA) and NINS (Japan), together
with NRC (Canada) and NSC and ASIAA (Taiwan) and KASI (Republic of
Korea), in cooperation with the Republic of Chile. The Joint ALMA
Observatory is operated by ESO, AUI/NRAO and NAOJ. MCW, LM and AS
acknowledge the support of the European Union through ERC grant number
279973. GMK is supported by the Royal Society as a Royal Society
University Research Fellow. AS is partially supported by funding from
the Center for Exoplanets and Habitable Worlds. The Center for
Exoplanets and Habitable Worlds is supported by the Pennsylvania State
University, the Eberly College of Science, and the Pennsylvania Space
Grant Consortium.

%%%%%%%%%%%%%%%%%%%%%%%%%%%%%%%%%%%%%%%%%%%%%%%%%%



\bibliographystyle{mnras}
\bibliography{SM_pformation} % if your bibtex file is called example.bib

\newcommand{\noop}[1]{}
\begin{thebibliography}{}
\makeatletter
\relax
\def\mn@urlcharsother{\let\do\@makeother \do\$\do\&\do\#\do\^\do\_\do\%\do\~}
\def\mn@doi{\begingroup\mn@urlcharsother \@ifnextchar [ {\mn@doi@}
  {\mn@doi@[]}}
\def\mn@doi@[#1]#2{\def\@tempa{#1}\ifx\@tempa\@empty \href
  {http://dx.doi.org/#2} {doi:#2}\else \href {http://dx.doi.org/#2} {#1}\fi
  \endgroup}
\def\mn@eprint#1#2{\mn@eprint@#1:#2::\@nil}
\def\mn@eprint@arXiv#1{\href {http://arxiv.org/abs/#1} {{\tt arXiv:#1}}}
\def\mn@eprint@dblp#1{\href {http://dblp.uni-trier.de/rec/bibtex/#1.xml}
  {dblp:#1}}
\def\mn@eprint@#1:#2:#3:#4\@nil{\def\@tempa {#1}\def\@tempb {#2}\def\@tempc
  {#3}\ifx \@tempc \@empty \let \@tempc \@tempb \let \@tempb \@tempa \fi \ifx
  \@tempb \@empty \def\@tempb {arXiv}\fi \@ifundefined
  {mn@eprint@\@tempb}{\@tempb:\@tempc}{\expandafter \expandafter \csname
  mn@eprint@\@tempb\endcsname \expandafter{\@tempc}}}

\bibitem[\protect\citeauthoryear{{Alibert} et~al.,}{{Alibert}
  et~al.}{2006}]{Alibert2006}
{Alibert} Y.,  et~al., 2006, \mn@doi [\aap] {10.1051/0004-6361:20065697}, \href
  {http://adsabs.harvard.edu/abs/2006A%26A...455L..25A} {455, L25}

\bibitem[\protect\citeauthoryear{{Benz} \& {Asphaug}}{{Benz} \&
  {Asphaug}}{1999}]{Benz1999}
{Benz} W.,  {Asphaug} E.,  1999, \mn@doi [\icarus] {10.1006/icar.1999.6204},
  \href {http://adsabs.harvard.edu/abs/1999Icar..142....5B} {142, 5}

\bibitem[\protect\citeauthoryear{{Biggs} et~al.,}{{Biggs}
  et~al.}{2011}]{Biggs2011}
{Biggs} A.~D.,  et~al., 2011, \mn@doi [\mnras]
  {10.1111/j.1365-2966.2010.18132.x}, \href
  {http://adsabs.harvard.edu/abs/2011MNRAS.413.2314B} {413, 2314}

\bibitem[\protect\citeauthoryear{Bohren \& Huffman}{Bohren \&
  Huffman}{1983}]{BohrenHuffman1983}
Bohren C.~F.,  Huffman D.,  1983, Absorption and scattering of light by small
  particles.
Wiley science paperback series, Wiley, \url
  {http://books.google.cl/books?id=S1RCZ8BjgN0C}

\bibitem[\protect\citeauthoryear{{Booth} et~al.,}{{Booth}
  et~al.}{2016}]{Booth2016}
{Booth} M.,  et~al., 2016, \mn@doi [\mnras] {10.1093/mnrasl/slw040}, \href
  {http://adsabs.harvard.edu/abs/2016MNRAS.tmpL..24B} {}

\bibitem[\protect\citeauthoryear{{Bryden} et~al.,}{{Bryden}
  et~al.}{2006}]{Bryden2006}
{Bryden} G.,  et~al., 2006, \mn@doi [\apj] {10.1086/498093}, \href
  {http://adsabs.harvard.edu/abs/2006ApJ...636.1098B} {636, 1098}

\bibitem[\protect\citeauthoryear{{Bryden} et~al.,}{{Bryden}
  et~al.}{2009}]{Bryden2009}
{Bryden} G.,  et~al., 2009, \mn@doi [\apj] {10.1088/0004-637X/705/2/1226},
  \href {http://adsabs.harvard.edu/abs/2009ApJ...705.1226B} {705, 1226}

\bibitem[\protect\citeauthoryear{{Carpenter} et~al.,}{{Carpenter}
  et~al.}{2009}]{Carpenter2009}
{Carpenter} J.~M.,  et~al., 2009, \mn@doi [\apjs]
  {10.1088/0067-0049/181/1/197}, \href
  {http://adsabs.harvard.edu/abs/2009ApJS..181..197C} {181, 197}

\bibitem[\protect\citeauthoryear{{Cataldi} et~al.,}{{Cataldi}
  et~al.}{2014}]{Cataldi2014}
{Cataldi} G.,  et~al., 2014, \mn@doi [\aap] {10.1051/0004-6361/201323126},
  \href {http://adsabs.harvard.edu/abs/2014A%26A...563A..66C} {563, A66}

\bibitem[\protect\citeauthoryear{{Cavanagh}, {Jenness}, {Economou}  \&
  {Currie}}{{Cavanagh} et~al.}{2008}]{Cavanagh2008}
{Cavanagh} B.,  {Jenness} T.,  {Economou} F.,   {Currie} M.~J.,  2008, \mn@doi
  [Astronomische Nachrichten] {10.1002/asna.200710944}, \href
  {http://adsabs.harvard.edu/abs/2008AN....329..295C} {329, 295}

\bibitem[\protect\citeauthoryear{{Chapin}, {Berry}, {Gibb}, {Jenness}, {Scott},
  {Tilanus}, {Economou}  \& {Holland}}{{Chapin} et~al.}{2013}]{Chapin2013}
{Chapin} E.~L.,  {Berry} D.~S.,  {Gibb} A.~G.,  {Jenness} T.,  {Scott} D.,
  {Tilanus} R.~P.~J.,  {Economou} F.,   {Holland} W.~S.,  2013, \mn@doi
  [\mnras] {10.1093/mnras/stt052}, \href
  {http://adsabs.harvard.edu/abs/2013MNRAS.430.2545C} {430, 2545}

\bibitem[\protect\citeauthoryear{{Chiang} \& {Laughlin}}{{Chiang} \&
  {Laughlin}}{2013}]{Chiang2013}
{Chiang} E.,  {Laughlin} G.,  2013, \mn@doi [\mnras] {10.1093/mnras/stt424},
  \href {http://adsabs.harvard.edu/abs/2013MNRAS.431.3444C} {431, 3444}

\bibitem[\protect\citeauthoryear{{Condon}, {Cotton}, {Greisen}, {Yin},
  {Perley}, {Taylor}  \& {Broderick}}{{Condon} et~al.}{1998}]{Condon1998}
{Condon} J.~J.,  {Cotton} W.~D.,  {Greisen} E.~W.,  {Yin} Q.~F.,  {Perley}
  R.~A.,  {Taylor} G.~B.,   {Broderick} J.~J.,  1998, \mn@doi [\aj]
  {10.1086/300337}, \href {http://adsabs.harvard.edu/abs/1998AJ....115.1693C}
  {115, 1693}

\bibitem[\protect\citeauthoryear{{Davis} \& {Ryan}}{{Davis} \&
  {Ryan}}{1990}]{Davis1990}
{Davis} D.~R.,  {Ryan} E.~V.,  1990, \mn@doi [\icarus]
  {10.1016/0019-1035(90)90012-X}, \href
  {http://adsabs.harvard.edu/abs/1990Icar...83..156D} {83, 156}

\bibitem[\protect\citeauthoryear{{Dent} et~al.,}{{Dent}
  et~al.}{2014}]{Dent2014}
{Dent} W.~R.~F.,  et~al., 2014, \mn@doi [Science] {10.1126/science.1248726},
  \href {http://adsabs.harvard.edu/abs/2014Sci...343.1490D} {343, 1490}

\bibitem[\protect\citeauthoryear{{Dohnanyi}}{{Dohnanyi}}{1969}]{Dohnanyi1969}
{Dohnanyi} J.~S.,  1969, \mn@doi [\jgr] {10.1029/JB074i010p02531}, \href
  {http://adsabs.harvard.edu/abs/1969JGR....74.2531D} {74, 2531}

\bibitem[\protect\citeauthoryear{{Dominik} \& {Decin}}{{Dominik} \&
  {Decin}}{2003}]{Dominik2003}
{Dominik} C.,  {Decin} G.,  2003, \mn@doi [\apj] {10.1086/379169}, \href
  {http://adsabs.harvard.edu/abs/2003ApJ...598..626D} {598, 626}

\bibitem[\protect\citeauthoryear{{Draine}}{{Draine}}{2003}]{Draine2003}
{Draine} B.~T.,  2003, \mn@doi [\apj] {10.1086/379118}, \href
  {http://adsabs.harvard.edu/abs/2003ApJ...598.1017D} {598, 1017}

\bibitem[\protect\citeauthoryear{{Dullemond}, {Juhasz}, {Pohl}, {Sereshti},
  {Shetty}, {Peters}, {Commercon}  \& {Flock}}{{Dullemond}
  et~al.}{2016}]{RADMC3D0.40}
{Dullemond} C.,  {Juhasz} A.,  {Pohl} A.,  {Sereshti} F.,  {Shetty} R.,
  {Peters} T.,  {Commercon} B.,   {Flock} M.,  2016, RADMC3D v0.40
  {http://www.ita.uni-heidelberg.de/ dullemond/software/radmc-3d/}

\bibitem[\protect\citeauthoryear{{Duncan} \& {Levison}}{{Duncan} \&
  {Levison}}{1997}]{Duncan1997}
{Duncan} M.~J.,  {Levison} H.~F.,  1997, \mn@doi [Science]
  {10.1126/science.276.5319.1670}, \href
  {http://adsabs.harvard.edu/abs/1997Sci...276.1670D} {276, 1670}

\bibitem[\protect\citeauthoryear{{Durda} \& {Dermott}}{{Durda} \&
  {Dermott}}{1997}]{Durda1997}
{Durda} D.~D.,  {Dermott} S.~F.,  1997, \mn@doi [\icarus]
  {10.1006/icar.1997.5803}, \href
  {http://adsabs.harvard.edu/abs/1997Icar..130..140D} {130, 140}

\bibitem[\protect\citeauthoryear{{Eiroa} et~al.,}{{Eiroa}
  et~al.}{2013}]{Eiroa2013}
{Eiroa} C.,  et~al., 2013, \mn@doi [\aap] {10.1051/0004-6361/201321050}, \href
  {http://adsabs.harvard.edu/abs/2013A%26A...555A..11E} {555, A11}

\bibitem[\protect\citeauthoryear{{Fontenla}, {Balasubramaniam}  \&
  {Harder}}{{Fontenla} et~al.}{2007}]{Fontenla2007}
{Fontenla} J.~M.,  {Balasubramaniam} K.~S.,   {Harder} J.,  2007, \mn@doi
  [\apj] {10.1086/520319}, \href
  {http://adsabs.harvard.edu/abs/2007ApJ...667.1243F} {667, 1243}

\bibitem[\protect\citeauthoryear{{Foreman-Mackey}, {Hogg}, {Lang}  \&
  {Goodman}}{{Foreman-Mackey} et~al.}{2013}]{emcee}
{Foreman-Mackey} D.,  {Hogg} D.~W.,  {Lang} D.,   {Goodman} J.,  2013, \mn@doi
  [PASP] {10.1086/670067}, 125, 306

\bibitem[\protect\citeauthoryear{Foreman-Mackey, Price-Whelan, Ryan, Emily,
  Smith, Barbary, Hogg  \& Brewer}{Foreman-Mackey et~al.}{2014}]{cornerplot}
Foreman-Mackey D.,  Price-Whelan A.,  Ryan G.,  Emily Smith M.,  Barbary K.,
  Hogg D.~W.,   Brewer B.~J.,  2014, triangle.py v0.1.1,
  \mn@doi{10.5281/zenodo.11020}, \url {http://dx.doi.org/10.5281/zenodo.11020}

\bibitem[\protect\citeauthoryear{{Fressin} et~al.,}{{Fressin}
  et~al.}{2013}]{Fressin2013}
{Fressin} F.,  et~al., 2013, \mn@doi [\apj] {10.1088/0004-637X/766/2/81}, \href
  {http://adsabs.harvard.edu/abs/2013ApJ...766...81F} {766, 81}

\bibitem[\protect\citeauthoryear{{Fujiwara}, {Cerroni}, {Davis}, {Ryan}  \& {di
  Martino}}{{Fujiwara} et~al.}{1989}]{Fujiwara1989}
{Fujiwara} A.,  {Cerroni} P.,  {Davis} D.,  {Ryan} E.,   {di Martino} M.,
  1989, in {Binzel} R.~P.,  {Gehrels} T.,   {Matthews} M.~S.,  eds, Asteroids
  II. pp 240--265

\bibitem[\protect\citeauthoryear{{Geiler} \& {Krivov}}{{Geiler} \&
  {Krivov}}{2017}]{Geiler2017}
{Geiler} F.,  {Krivov} A.,  2017, preprint, \href
  {http://adsabs.harvard.edu/abs/2017arXiv170205966G} {} (\mn@eprint {arXiv}
  {1702.05966})

\bibitem[\protect\citeauthoryear{Goodman \& Weare}{Goodman \&
  Weare}{2010}]{GoodmanWeare2010}
Goodman J.,  Weare J.,  2010, \mn@doi [Commun. Appl. Math. Comput. Sci.]
  {10.2140/camcos.2010.5.65}, 5, 65

\bibitem[\protect\citeauthoryear{{Greaves}, {Holland}, {Jayawardhana}, {Wyatt}
  \& {Dent}}{{Greaves} et~al.}{2004}]{Greaves2004}
{Greaves} J.~S.,  {Holland} W.~S.,  {Jayawardhana} R.,  {Wyatt} M.~C.,   {Dent}
  W.~R.~F.,  2004, \mn@doi [\mnras] {10.1111/j.1365-2966.2004.07440.x}, \href
  {http://adsabs.harvard.edu/abs/2004MNRAS.348.1097G} {348, 1097}

\bibitem[\protect\citeauthoryear{{Hansen} \& {Murray}}{{Hansen} \&
  {Murray}}{2012}]{Hansen2012}
{Hansen} B.~M.~S.,  {Murray} N.,  2012, \mn@doi [\apj]
  {10.1088/0004-637X/751/2/158}, \href
  {http://adsabs.harvard.edu/abs/2012ApJ...751..158H} {751, 158}

\bibitem[\protect\citeauthoryear{{Hayashi}}{{Hayashi}}{1981}]{Hayashi1981}
{Hayashi} C.,  1981, \mn@doi [Progress of Theoretical Physics Supplement]
  {10.1143/PTPS.70.35}, \href
  {http://adsabs.harvard.edu/abs/1981PThPS..70...35H} {70, 35}

\bibitem[\protect\citeauthoryear{{Hillenbrand} et~al.,}{{Hillenbrand}
  et~al.}{2008}]{Hillenbrand2008}
{Hillenbrand} L.~A.,  et~al., 2008, \mn@doi [\apj] {10.1086/529027}, \href
  {http://adsabs.harvard.edu/abs/2008ApJ...677..630H} {677, 630}

\bibitem[\protect\citeauthoryear{{Holland} et~al.,}{{Holland}
  et~al.}{2013}]{Holland2013}
{Holland} W.~S.,  et~al., 2013, \mn@doi [\mnras] {10.1093/mnras/sts612}, \href
  {http://adsabs.harvard.edu/abs/2013MNRAS.430.2513H} {430, 2513}

\bibitem[\protect\citeauthoryear{{Howard} et~al.,}{{Howard}
  et~al.}{2010}]{Howard2010}
{Howard} A.~W.,  et~al., 2010, \mn@doi [Science] {10.1126/science.1194854},
  \href {http://adsabs.harvard.edu/abs/2010Sci...330..653H} {330, 653}

\bibitem[\protect\citeauthoryear{{Ida} \& {Lin}}{{Ida} \&
  {Lin}}{2010}]{IdaLin2010}
{Ida} S.,  {Lin} D.~N.~C.,  2010, \mn@doi [\apj] {10.1088/0004-637X/719/1/810},
  \href {http://adsabs.harvard.edu/abs/2010ApJ...719..810I} {719, 810}

\bibitem[\protect\citeauthoryear{{Ikarashi} et~al.,}{{Ikarashi}
  et~al.}{2015}]{Ikarashi2015}
{Ikarashi} S.,  et~al., 2015, \mn@doi [\apj] {10.1088/0004-637X/810/2/133},
  \href {http://adsabs.harvard.edu/abs/2015ApJ...810..133I} {810, 133}

\bibitem[\protect\citeauthoryear{{Kalas} et~al.,}{{Kalas}
  et~al.}{2008}]{Kalas2008}
{Kalas} P.,  et~al., 2008, \mn@doi [Science] {10.1126/science.1166609}, \href
  {http://adsabs.harvard.edu/abs/2008Sci...322.1345K} {322, 1345}

\bibitem[\protect\citeauthoryear{{Kennedy} \& {Kenyon}}{{Kennedy} \&
  {Kenyon}}{2008}]{Kennedy2008hotsuperearth}
{Kennedy} G.~M.,  {Kenyon} S.~J.,  2008, \mn@doi [\apj] {10.1086/589436}, \href
  {http://adsabs.harvard.edu/abs/2008ApJ...682.1264K} {682, 1264}

\bibitem[\protect\citeauthoryear{{Kennedy} et~al.,}{{Kennedy}
  et~al.}{2015}]{Kennedy2015superearths}
{Kennedy} G.~M.,  et~al., 2015, \mn@doi [\mnras] {10.1093/mnras/stv511}, \href
  {http://adsabs.harvard.edu/abs/2015MNRAS.449.3121K} {449, 3121}

\bibitem[\protect\citeauthoryear{{Kenyon} \& {Bromley}}{{Kenyon} \&
  {Bromley}}{2008}]{Kenyon2008}
{Kenyon} S.~J.,  {Bromley} B.~C.,  2008, \mn@doi [\apjs] {10.1086/591794},
  \href {http://adsabs.harvard.edu/abs/2008ApJS..179..451K} {179, 451}

\bibitem[\protect\citeauthoryear{{Kral}, {Wyatt}, {Carswell}, {Pringle},
  {Matr{\`a}}  \& {Juh{\'a}sz}}{{Kral} et~al.}{2016}]{Kral2016}
{Kral} Q.,  {Wyatt} M.,  {Carswell} R.~F.,  {Pringle} J.~E.,  {Matr{\`a}} L.,
  {Juh{\'a}sz} A.,  2016, \mn@doi [\mnras] {10.1093/mnras/stw1361}, \href
  {http://adsabs.harvard.edu/abs/2016MNRAS.461..845K} {461, 845}

\bibitem[\protect\citeauthoryear{{Krivov}, {L{\"o}hne}  \& {Srem{\v
  c}evi{\'c}}}{{Krivov} et~al.}{2006}]{Krivov2006}
{Krivov} A.~V.,  {L{\"o}hne} T.,   {Srem{\v c}evi{\'c}} M.,  2006, \mn@doi
  [\aap] {10.1051/0004-6361:20064907}, \href
  {http://adsabs.harvard.edu/abs/2006A%26A...455..509K} {455, 509}

\bibitem[\protect\citeauthoryear{{Kuchner}}{{Kuchner}}{2004}]{Kuchner2004}
{Kuchner} M.~J.,  2004, \mn@doi [\apj] {10.1086/422577}, \href
  {http://adsabs.harvard.edu/abs/2004ApJ...612.1147K} {612, 1147}

\bibitem[\protect\citeauthoryear{{Kurucz}}{{Kurucz}}{1979}]{Kurucz1979}
{Kurucz} R.~L.,  1979, \mn@doi [\apjs] {10.1086/190589}, \href
  {http://cdsads.u-strasbg.fr/abs/1979ApJS...40....1K} {40, 1}

\bibitem[\protect\citeauthoryear{{Lagrange} et~al.,}{{Lagrange}
  et~al.}{2009}]{Lagrange2009betapicb}
{Lagrange} A.-M.,  et~al., 2009, \mn@doi [\aap] {10.1051/0004-6361:200811325},
  \href {http://adsabs.harvard.edu/abs/2009A%26A...493L..21L} {493, L21}

\bibitem[\protect\citeauthoryear{{Li} \& {Greenberg}}{{Li} \&
  {Greenberg}}{1998}]{LiGreenberg1998}
{Li} A.,  {Greenberg} J.~M.,  1998, \aap, \href
  {http://adsabs.harvard.edu/abs/1998A%26A...331..291L} {331, 291}

\bibitem[\protect\citeauthoryear{{Lissauer}}{{Lissauer}}{1993}]{Lissauer1993}
{Lissauer} J.~J.,  1993, \mn@doi [\araa] {10.1146/annurev.aa.31.090193.001021},
  \href {http://adsabs.harvard.edu/abs/1993ARA%26A..31..129L} {31, 129}

\bibitem[\protect\citeauthoryear{{L{\"o}hne}, {Krivov}  \&
  {Rodmann}}{{L{\"o}hne} et~al.}{2008}]{Lohne2008}
{L{\"o}hne} T.,  {Krivov} A.~V.,   {Rodmann} J.,  2008, \mn@doi [\apj]
  {10.1086/524840}, \href {http://adsabs.harvard.edu/abs/2008ApJ...673.1123L}
  {673, 1123}

\bibitem[\protect\citeauthoryear{{Loukitcheva}, {Solanki}, {Carlsson}  \&
  {Stein}}{{Loukitcheva} et~al.}{2004}]{Loukitcheva2004}
{Loukitcheva} M.,  {Solanki} S.~K.,  {Carlsson} M.,   {Stein} R.~F.,  2004,
  \mn@doi [\aap] {10.1051/0004-6361:20034159}, \href
  {http://adsabs.harvard.edu/abs/2004A%26A...419..747L} {419, 747}

\bibitem[\protect\citeauthoryear{{Marino}, {Casassus}, {Perez}, {Lyra},
  {Roman}, {Avenhaus}, {Wright}  \& {Maddison}}{{Marino}
  et~al.}{2015}]{Marino2015mwc}
{Marino} S.,  {Casassus} S.,  {Perez} S.,  {Lyra} W.,  {Roman} P.~E.,
  {Avenhaus} H.,  {Wright} C.~M.,   {Maddison} S.~T.,  2015, \mn@doi [\apj]
  {10.1088/0004-637X/813/1/76}, \href
  {http://adsabs.harvard.edu/abs/2015ApJ...813...76M} {813, 76}

\bibitem[\protect\citeauthoryear{{Marino} et~al.,}{{Marino}
  et~al.}{2016b}]{Marino2017}
{Marino} S.,  et~al., 2016b, preprint, \href
  {http://adsabs.harvard.edu/abs/2016arXiv161101168M} {} (\mn@eprint {arXiv}
  {1611.01168})

\bibitem[\protect\citeauthoryear{{Marino} et~al.,}{{Marino}
  et~al.}{2016a}]{Marino2016}
{Marino} S.,  et~al., 2016a, \mn@doi [\mnras] {10.1093/mnras/stw1216}, \href
  {http://adsabs.harvard.edu/abs/2016MNRAS.tmp..891M} {}

\bibitem[\protect\citeauthoryear{{Marois}, {Macintosh}, {Barman}, {Zuckerman},
  {Song}, {Patience}, {Lafreni{\`e}re}  \& {Doyon}}{{Marois}
  et~al.}{2008}]{Marois2008}
{Marois} C.,  {Macintosh} B.,  {Barman} T.,  {Zuckerman} B.,  {Song} I.,
  {Patience} J.,  {Lafreni{\`e}re} D.,   {Doyon} R.,  2008, \mn@doi [Science]
  {10.1126/science.1166585}, \href
  {http://adsabs.harvard.edu/abs/2008Sci...322.1348M} {322, 1348}

\bibitem[\protect\citeauthoryear{{Marois}, {Zuckerman}, {Konopacky},
  {Macintosh}  \& {Barman}}{{Marois} et~al.}{2010}]{Marois2010}
{Marois} C.,  {Zuckerman} B.,  {Konopacky} Q.~M.,  {Macintosh} B.,   {Barman}
  T.,  2010, \mn@doi [\nat] {10.1038/nature09684}, \href
  {http://adsabs.harvard.edu/abs/2010Natur.468.1080M} {468, 1080}

\bibitem[\protect\citeauthoryear{{Matr{\`a}}, {Pani{\'c}}, {Wyatt}  \&
  {Dent}}{{Matr{\`a}} et~al.}{2015}]{Matra2015}
{Matr{\`a}} L.,  {Pani{\'c}} O.,  {Wyatt} M.~C.,   {Dent} W.~R.~F.,  2015,
  \mn@doi [\mnras] {10.1093/mnras/stu2619}, \href
  {http://adsabs.harvard.edu/abs/2015MNRAS.447.3936M} {447, 3936}

\bibitem[\protect\citeauthoryear{{Matr{\`a}} et~al.,}{{Matr{\`a}}
  et~al.}{2017}]{Matra2017betapic}
{Matr{\`a}} L.,  et~al., 2017, \mn@doi [\mnras] {10.1093/mnras/stw2415}, \href
  {http://adsabs.harvard.edu/abs/2017MNRAS.464.1415M} {464, 1415}

\bibitem[\protect\citeauthoryear{{Matr{\`a}} et~al.}{{Matr{\`a}}
  et~al.}{\noop{1001}submitted}]{Matra2017Fomalhaut}
{Matr{\`a}} L.,  et~al., \noop{1001}submitted

\bibitem[\protect\citeauthoryear{{Matthews}, {Krivov}, {Wyatt}, {Bryden}  \&
  {Eiroa}}{{Matthews} et~al.}{2014a}]{Matthews2014pp6}
{Matthews} B.~C.,  {Krivov} A.~V.,  {Wyatt} M.~C.,  {Bryden} G.,   {Eiroa} C.,
  2014a, \mn@doi [Protostars and Planets VI]
  {10.2458/azu_uapress_9780816531240-ch023}, \href
  {http://adsabs.harvard.edu/abs/2014prpl.conf..521M} {pp 521--544}

\bibitem[\protect\citeauthoryear{{Matthews}, {Kennedy}, {Sibthorpe}, {Booth},
  {Wyatt}, {Broekhoven-Fiene}, {Macintosh}  \& {Marois}}{{Matthews}
  et~al.}{2014b}]{Matthews2014hr8799}
{Matthews} B.,  {Kennedy} G.,  {Sibthorpe} B.,  {Booth} M.,  {Wyatt} M.,
  {Broekhoven-Fiene} H.,  {Macintosh} B.,   {Marois} C.,  2014b, \mn@doi [\apj]
  {10.1088/0004-637X/780/1/97}, \href
  {http://adsabs.harvard.edu/abs/2014ApJ...780...97M} {780, 97}

\bibitem[\protect\citeauthoryear{{Matthews} et~al.,}{{Matthews}
  et~al.}{2015}]{Matthews2015}
{Matthews} B.~C.,  et~al., 2015, \mn@doi [\apj] {10.1088/0004-637X/811/2/100},
  \href {http://adsabs.harvard.edu/abs/2015ApJ...811..100M} {811, 100}

\bibitem[\protect\citeauthoryear{{Mayor} et~al.,}{{Mayor}
  et~al.}{2011}]{Mayor2011}
{Mayor} M.,  et~al., 2011, preprint, \href
  {http://adsabs.harvard.edu/abs/2011arXiv1109.2497M} {} (\mn@eprint {arXiv}
  {1109.2497})

\bibitem[\protect\citeauthoryear{{McMullin}, {Waters}, {Schiebel}, {Young}  \&
  {Golap}}{{McMullin} et~al.}{2007}]{McMullin2007}
{McMullin} J.~P.,  {Waters} B.,  {Schiebel} D.,  {Young} W.,   {Golap} K.,
  2007, in {Shaw} R.~A.,  {Hill} F.,   {Bell} D.~J.,  eds,  Astronomical
  Society of the Pacific Conference Series Vol. 376, Astronomical Data Analysis
  Software and Systems XVI. p.~127

\bibitem[\protect\citeauthoryear{{Montesinos} et~al.,}{{Montesinos}
  et~al.}{2016}]{Montesinos2016}
{Montesinos} B.,  et~al., 2016, \mn@doi [\aap] {10.1051/0004-6361/201628329},
  \href {http://adsabs.harvard.edu/abs/2016A%26A...593A..51M} {593, A51}

\bibitem[\protect\citeauthoryear{{Mo{\'o}r} et~al.,}{{Mo{\'o}r}
  et~al.}{2015a}]{Moor2015herschel}
{Mo{\'o}r} A.,  et~al., 2015a, \mn@doi [\mnras] {10.1093/mnras/stu2442}, \href
  {http://adsabs.harvard.edu/abs/2015MNRAS.447..577M} {447, 577}

\bibitem[\protect\citeauthoryear{{Mo{\'o}r} et~al.,}{{Mo{\'o}r}
  et~al.}{2015b}]{Moor2015gas}
{Mo{\'o}r} A.,  et~al., 2015b, \mn@doi [\apj] {10.1088/0004-637X/814/1/42},
  \href {http://adsabs.harvard.edu/abs/2015ApJ...814...42M} {814, 42}

\bibitem[\protect\citeauthoryear{{Moro-Mart{\'{\i}}n}
  et~al.,}{{Moro-Mart{\'{\i}}n} et~al.}{2007}]{Moro-Martin2007}
{Moro-Mart{\'{\i}}n} A.,  et~al., 2007, \mn@doi [\apj] {10.1086/511746}, \href
  {http://adsabs.harvard.edu/abs/2007ApJ...658.1312M} {658, 1312}

\bibitem[\protect\citeauthoryear{{Moro-Mart{\'{\i}}n}
  et~al.,}{{Moro-Mart{\'{\i}}n} et~al.}{2015}]{Moro-Martin2015}
{Moro-Mart{\'{\i}}n} A.,  et~al., 2015, \mn@doi [\apj]
  {10.1088/0004-637X/801/2/143}, \href
  {http://adsabs.harvard.edu/abs/2015ApJ...801..143M} {801, 143}

\bibitem[\protect\citeauthoryear{{Mumma} \& {Charnley}}{{Mumma} \&
  {Charnley}}{2011}]{Mumma2011}
{Mumma} M.~J.,  {Charnley} S.~B.,  2011, \mn@doi [\araa]
  {10.1146/annurev-astro-081309-130811}, \href
  {http://adsabs.harvard.edu/abs/2011ARA%26A..49..471M} {49, 471}

\bibitem[\protect\citeauthoryear{{Mustill} \& {Wyatt}}{{Mustill} \&
  {Wyatt}}{2009}]{Mustill2009}
{Mustill} A.~J.,  {Wyatt} M.~C.,  2009, \mn@doi [\mnras]
  {10.1111/j.1365-2966.2009.15360.x}, \href
  {http://adsabs.harvard.edu/abs/2009MNRAS.399.1403M} {399, 1403}

\bibitem[\protect\citeauthoryear{{O'Brien} \& {Greenberg}}{{O'Brien} \&
  {Greenberg}}{2003}]{OBrien2003}
{O'Brien} D.~P.,  {Greenberg} R.,  2003, \mn@doi [\icarus]
  {10.1016/S0019-1035(03)00145-3}, \href
  {http://adsabs.harvard.edu/abs/2003Icar..164..334O} {164, 334}

\bibitem[\protect\citeauthoryear{{Pan}, {Nesvold}  \& {Kuchner}}{{Pan}
  et~al.}{2016}]{Pan2016}
{Pan} M.,  {Nesvold} E.~R.,   {Kuchner} M.~J.,  2016, preprint, \href
  {http://adsabs.harvard.edu/abs/2016arXiv160706798P} {} (\mn@eprint {arXiv}
  {1607.06798})

\bibitem[\protect\citeauthoryear{{Pani{\'c}} et~al.,}{{Pani{\'c}}
  et~al.}{2013}]{Panic2013}
{Pani{\'c}} O.,  et~al., 2013, \mn@doi [\mnras] {10.1093/mnras/stt1293}, \href
  {http://adsabs.harvard.edu/abs/2013MNRAS.435.1037P} {435, 1037}

\bibitem[\protect\citeauthoryear{{Payne}, {Ford}, {Wyatt}  \& {Booth}}{{Payne}
  et~al.}{2009}]{Payne2009}
{Payne} M.~J.,  {Ford} E.~B.,  {Wyatt} M.~C.,   {Booth} M.,  2009, \mn@doi
  [\mnras] {10.1111/j.1365-2966.2008.14338.x}, \href
  {http://adsabs.harvard.edu/abs/2009MNRAS.393.1219P} {393, 1219}

\bibitem[\protect\citeauthoryear{{Pearce} \& {Wyatt}}{{Pearce} \&
  {Wyatt}}{2014}]{Pearce2014}
{Pearce} T.~D.,  {Wyatt} M.~C.,  2014, \mn@doi [\mnras]
  {10.1093/mnras/stu1302}, \href
  {http://adsabs.harvard.edu/abs/2014MNRAS.443.2541P} {443, 2541}

\bibitem[\protect\citeauthoryear{{Raymond}, {Quinn}  \& {Lunine}}{{Raymond}
  et~al.}{2005}]{Raymond2005}
{Raymond} S.~N.,  {Quinn} T.,   {Lunine} J.~I.,  2005, \mn@doi [\apj]
  {10.1086/433179}, \href {http://adsabs.harvard.edu/abs/2005ApJ...632..670R}
  {632, 670}

\bibitem[\protect\citeauthoryear{{Raymond} et~al.,}{{Raymond}
  et~al.}{2011}]{Raymond2011}
{Raymond} S.~N.,  et~al., 2011, \mn@doi [\aap] {10.1051/0004-6361/201116456},
  \href {http://adsabs.harvard.edu/abs/2011A%26A...530A..62R} {530, A62}

\bibitem[\protect\citeauthoryear{{Read} \& {Wyatt}}{{Read} \&
  {Wyatt}}{2016}]{Read2016}
{Read} M.~J.,  {Wyatt} M.~C.,  2016, \mn@doi [\mnras] {10.1093/mnras/stv2968},
  \href {http://adsabs.harvard.edu/abs/2016MNRAS.457..465R} {457, 465}

\bibitem[\protect\citeauthoryear{{Roberge}, {Feldman}, {Lagrange},
  {Vidal-Madjar}, {Ferlet}, {Jolly}, {Lemaire}  \& {Rostas}}{{Roberge}
  et~al.}{2000}]{Roberge2000}
{Roberge} A.,  {Feldman} P.~D.,  {Lagrange} A.~M.,  {Vidal-Madjar} A.,
  {Ferlet} R.,  {Jolly} A.,  {Lemaire} J.~L.,   {Rostas} F.,  2000, \mn@doi
  [\apj] {10.1086/309157}, \href
  {http://adsabs.harvard.edu/abs/2000ApJ...538..904R} {538, 904}

\bibitem[\protect\citeauthoryear{{Ryan}, {Hartmann}  \& {Davis}}{{Ryan}
  et~al.}{1991}]{Ryan1991}
{Ryan} E.~V.,  {Hartmann} W.~K.,   {Davis} D.~R.,  1991, \mn@doi [\icarus]
  {10.1016/0019-1035(91)90228-L}, \href
  {http://adsabs.harvard.edu/abs/1991Icar...94..283R} {94, 283}

\bibitem[\protect\citeauthoryear{{Sch{\"u}ppler}, {Krivov}, {L{\"o}hne},
  {Booth}, {Kirchschlager}  \& {Wolf}}{{Sch{\"u}ppler}
  et~al.}{2016}]{Schuppler2016}
{Sch{\"u}ppler} C.,  {Krivov} A.~V.,  {L{\"o}hne} T.,  {Booth} M.,
  {Kirchschlager} F.,   {Wolf} S.,  2016, \mn@doi [\mnras]
  {10.1093/mnras/stw1456}, \href
  {http://adsabs.harvard.edu/abs/2016MNRAS.461.2146S} {461, 2146}

\bibitem[\protect\citeauthoryear{{Shannon} \& {Wu}}{{Shannon} \&
  {Wu}}{2011}]{Shannon2011}
{Shannon} A.,  {Wu} Y.,  2011, \mn@doi [\apj] {10.1088/0004-637X/739/1/36},
  \href {http://adsabs.harvard.edu/abs/2011ApJ...739...36S} {739, 36}

\bibitem[\protect\citeauthoryear{{Simpson} et~al.,}{{Simpson}
  et~al.}{2015}]{Simpson2015}
{Simpson} J.~M.,  et~al., 2015, \mn@doi [\apj] {10.1088/0004-637X/799/1/81},
  \href {http://adsabs.harvard.edu/abs/2015ApJ...799...81S} {799, 81}

\bibitem[\protect\citeauthoryear{{Smail}, {Ivison}  \& {Blain}}{{Smail}
  et~al.}{1997}]{Smail1997}
{Smail} I.,  {Ivison} R.~J.,   {Blain} A.~W.,  1997, \mn@doi [\apjl]
  {10.1086/311017}, \href {http://adsabs.harvard.edu/abs/1997ApJ...490L...5S}
  {490, L5}

\bibitem[\protect\citeauthoryear{{Smith} \& {Terrile}}{{Smith} \&
  {Terrile}}{1984}]{Smith1984betapic}
{Smith} B.~A.,  {Terrile} R.~J.,  1984, \mn@doi [Science]
  {10.1126/science.226.4681.1421}, \href
  {http://adsabs.harvard.edu/abs/1984Sci...226.1421S} {226, 1421}

\bibitem[\protect\citeauthoryear{{Sousa} et~al.,}{{Sousa}
  et~al.}{2008}]{Sousa2008}
{Sousa} S.~G.,  et~al., 2008, \mn@doi [\aap] {10.1051/0004-6361:200809698},
  \href {http://adsabs.harvard.edu/abs/2008A%26A...487..373S} {487, 373}

\bibitem[\protect\citeauthoryear{{Stewart} \& {Leinhardt}}{{Stewart} \&
  {Leinhardt}}{2009}]{Stewart2009}
{Stewart} S.~T.,  {Leinhardt} Z.~M.,  2009, \mn@doi [\apjl]
  {10.1088/0004-637X/691/2/L133}, \href
  {http://adsabs.harvard.edu/abs/2009ApJ...691L.133S} {691, L133}

\bibitem[\protect\citeauthoryear{{Su} et~al.,}{{Su} et~al.}{2006}]{Su2006}
{Su} K.~Y.~L.,  et~al., 2006, \mn@doi [\apj] {10.1086/508649}, \href
  {http://adsabs.harvard.edu/abs/2006ApJ...653..675S} {653, 675}

\bibitem[\protect\citeauthoryear{{Terquem} \& {Papaloizou}}{{Terquem} \&
  {Papaloizou}}{2007}]{Terquem2007}
{Terquem} C.,  {Papaloizou} J.~C.~B.,  2007, \mn@doi [\apj] {10.1086/509497},
  \href {http://adsabs.harvard.edu/abs/2007ApJ...654.1110T} {654, 1110}

\bibitem[\protect\citeauthoryear{{Th{\'e}bault} \& {Augereau}}{{Th{\'e}bault}
  \& {Augereau}}{2007}]{Thebault2007}
{Th{\'e}bault} P.,  {Augereau} J.-C.,  2007, \mn@doi [\aap]
  {10.1051/0004-6361:20077709}, \href
  {http://adsabs.harvard.edu/abs/2007A%26A...472..169T} {472, 169}

\bibitem[\protect\citeauthoryear{{Thureau} et~al.,}{{Thureau}
  et~al.}{2014}]{Thureau2014}
{Thureau} N.~D.,  et~al., 2014, \mn@doi [\mnras] {10.1093/mnras/stu1864}, \href
  {http://adsabs.harvard.edu/abs/2014MNRAS.445.2558T} {445, 2558}

\bibitem[\protect\citeauthoryear{{Vican}}{{Vican}}{2012}]{Vican2012}
{Vican} L.,  2012, \mn@doi [\aj] {10.1088/0004-6256/143/6/135}, \href
  {http://adsabs.harvard.edu/abs/2012AJ....143..135V} {143, 135}

\bibitem[\protect\citeauthoryear{{Vogt} et~al.,}{{Vogt}
  et~al.}{2010}]{Vogt2010}
{Vogt} S.~S.,  et~al., 2010, \mn@doi [\apj] {10.1088/0004-637X/708/2/1366},
  \href {http://adsabs.harvard.edu/abs/2010ApJ...708.1366V} {708, 1366}

\bibitem[\protect\citeauthoryear{{Weidenschilling}}{{Weidenschilling}}{1977a}]{Widenschilling1977mmsn}
{Weidenschilling} S.~J.,  1977a, \mn@doi [\apss] {10.1007/BF00642464}, \href
  {http://adsabs.harvard.edu/abs/1977Ap%26SS..51..153W} {51, 153}

\bibitem[\protect\citeauthoryear{{Weidenschilling}}{{Weidenschilling}}{1977b}]{Weidenschilling1977drag}
{Weidenschilling} S.~J.,  1977b, \mnras, \href
  {http://adsabs.harvard.edu/abs/1977MNRAS.180...57W} {180, 57}

\bibitem[\protect\citeauthoryear{{Wetherill} \& {Stewart}}{{Wetherill} \&
  {Stewart}}{1993}]{Wetherill1993}
{Wetherill} G.~W.,  {Stewart} G.~R.,  1993, \mn@doi [\icarus]
  {10.1006/icar.1993.1166}, \href
  {http://adsabs.harvard.edu/abs/1993Icar..106..190W} {106, 190}

\bibitem[\protect\citeauthoryear{{Whipple}}{{Whipple}}{1973}]{Whipple1973}
{Whipple} F.~L.,  1973, NASA Special Publication, \href
  {http://adsabs.harvard.edu/abs/1973NASSP.319..355W} {319, 355}

\bibitem[\protect\citeauthoryear{{Wright}, {Drake}, {Mamajek}  \&
  {Henry}}{{Wright} et~al.}{2011}]{Wright2011}
{Wright} N.~J.,  {Drake} J.~J.,  {Mamajek} E.~E.,   {Henry} G.~W.,  2011,
  \mn@doi [\apj] {10.1088/0004-637X/743/1/48}, \href
  {http://adsabs.harvard.edu/abs/2011ApJ...743...48W} {743, 48}

\bibitem[\protect\citeauthoryear{{Wyatt}}{{Wyatt}}{2006}]{Wyatt2006}
{Wyatt} M.~C.,  2006, \mn@doi [\apj] {10.1086/499487}, \href
  {http://adsabs.harvard.edu/abs/2006ApJ...639.1153W} {639, 1153}

\bibitem[\protect\citeauthoryear{{Wyatt}, {Dermott}, {Telesco}, {Fisher},
  {Grogan}, {Holmes}  \& {Pi{\~n}a}}{{Wyatt} et~al.}{1999}]{Wyatt1999}
{Wyatt} M.~C.,  {Dermott} S.~F.,  {Telesco} C.~M.,  {Fisher} R.~S.,  {Grogan}
  K.,  {Holmes} E.~K.,   {Pi{\~n}a} R.~K.,  1999, \mn@doi [\apj]
  {10.1086/308093}, \href {http://adsabs.harvard.edu/abs/1999ApJ...527..918W}
  {527, 918}

\bibitem[\protect\citeauthoryear{{Wyatt}, {Smith}, {Su}, {Rieke}, {Greaves},
  {Beichman}  \& {Bryden}}{{Wyatt} et~al.}{2007}]{Wyatt2007collisionalcascade}
{Wyatt} M.~C.,  {Smith} R.,  {Su} K.~Y.~L.,  {Rieke} G.~H.,  {Greaves} J.~S.,
  {Beichman} C.~A.,   {Bryden} G.,  2007, \mn@doi [\apj] {10.1086/518404},
  \href {http://adsabs.harvard.edu/abs/2007ApJ...663..365W} {663, 365}

\bibitem[\protect\citeauthoryear{{Wyatt}, {Booth}, {Payne}  \&
  {Churcher}}{{Wyatt} et~al.}{2010}]{Wyatt2010}
{Wyatt} M.~C.,  {Booth} M.,  {Payne} M.~J.,   {Churcher} L.~J.,  2010, \mn@doi
  [\mnras] {10.1111/j.1365-2966.2009.15930.x}, \href
  {http://adsabs.harvard.edu/abs/2010MNRAS.402..657W} {402, 657}

\bibitem[\protect\citeauthoryear{{Wyatt}, {Clarke}  \& {Booth}}{{Wyatt}
  et~al.}{2011}]{Wyatt2011}
{Wyatt} M.~C.,  {Clarke} C.~J.,   {Booth} M.,  2011, \mn@doi [Celestial
  Mechanics and Dynamical Astronomy] {10.1007/s10569-011-9345-3}, \href
  {http://adsabs.harvard.edu/abs/2011CeMDA.111....1W} {111, 1}

\bibitem[\protect\citeauthoryear{{Wyatt} et~al.,}{{Wyatt}
  et~al.}{2012}]{Wyatt2012}
{Wyatt} M.~C.,  et~al., 2012, \mn@doi [\mnras]
  {10.1111/j.1365-2966.2012.21298.x}, \href
  {http://adsabs.harvard.edu/abs/2012MNRAS.424.1206W} {424, 1206}

\bibitem[\protect\citeauthoryear{{Zuckerman}, {Forveille}  \&
  {Kastner}}{{Zuckerman} et~al.}{1995}]{Zuckerman1995}
{Zuckerman} B.,  {Forveille} T.,   {Kastner} J.~H.,  1995, \mn@doi [\nat]
  {10.1038/373494a0}, \href {http://adsabs.harvard.edu/abs/1995Natur.373..494Z}
  {373, 494}

\bibitem[\protect\citeauthoryear{{van Leeuwen}}{{van
  Leeuwen}}{2007}]{vanLeeuwen2007}
{van Leeuwen} F.,  2007, \mn@doi [\aap] {10.1051/0004-6361:20078357}, \href
  {http://cdsads.u-strasbg.fr/abs/2007A%26A...474..653V} {474, 653}

\makeatother
\end{thebibliography}

%%%%%%%%%%%%%%%%%%%%%%%%%%%%%%%%%%%%%%%%%%%%%%%%%%

%%%%%%%%%%%%%%%%% APPENDICES %%%%%%%%%%%%%%%%%%%%%


%% \section{Some extra material}

%% If you want to present additional material which would interrupt the flow of the main paper,
%% it can be placed in an Appendix which appears after the list of references.

%%%%%%%%%%%%%%%%%%%%%%%%%%%%%%%%%%%%%%%%%%%%%%%%%%

% Don't change these lines
\bsp	% typesetting comment
\label{lastpage}
\end{document}